\documentclass[pre,twocolumn,superscriptaddress]{revtex4}
\usepackage{graphicx,subfigure}
\usepackage{amsthm}
\usepackage{amsmath}
\usepackage{color,colordvi}
\usepackage[skip=2pt,font=small]{caption}

\begin{document}
\bibliographystyle{revtex4}


\title{Electron power absorption dynamics in a low pressure radio frequency driven capacitively coupled discharge in oxygen}



\author{A. Proto}

\affiliation{Science Institute, University of Iceland,
                Dunhaga 3, IS-107 Reykjavik, Iceland}

\author{J. T. Gudmundsson}
\email[]{tumi@hi.is}

\affiliation{Science Institute, University of Iceland,
                Dunhaga 3, IS-107 Reykjavik, Iceland}

\affiliation{Department of Space and Plasma Physics, School of Electrical Engineering and Computer Science, 
KTH  Royal Institute of Technology, SE-100 44, Stockholm, Sweden}






\date{\today}

\begin{abstract}
We use the one-dimensional object-oriented particle-in-cell Monte Carlo collision code oopd1 to
explore the properties and the origins of both the electric field and electron power absorption within
the plasma bulk for a capacitively coupled oxygen discharge operated at 10 and 100 mTorr for 45
mm of gap distance. The properties of the electric field at three different time slices as well as time
averaged have been explored considering the moments of the Boltzmann equation. The electron power absorption is distinctly different at these operating pressures.  The most relevant contributions to the electric field at different time steps come from the pressure terms, the ambipolar and the electron temperature gradient terms, along with the ohmic term. The same applies for the electron power absorption. At both 10 and 100 mTorr the relative ohmic contribution to the electron power absorption remains roughly the same, while the ambipolar term contributes to power absorption and the temperature gradient term to electron cooling at 100 mTorr, and the opposite applies at 10 mTorr. At 100 mTorr the discharge is weekly electronegative, and electron power absorption is mainly due to sheath expansion while at 10 mTorr it is strongly
electronegative, and the electron power absorption occurs mainly within the electronegative core
and drift-ambipolar mode dominates. The agreement between the calculated values and the simulations is good for both the electric field and the electron power absorption within the plasma bulk and in the collapsed sheath region for all the cases considered.
\end{abstract}

\maketitle

\section{Introduction}\label{Intro}
\abovedisplayskip=10pt
\belowdisplayskip=10pt
The low pressure radio frequency (rf) driven capacitively coupled discharge has been applied in integrated circuit manufacturing for a few decades. The capacitively coupled discharge consists of two parallel electrodes, typically with a radius of few tens of cm, separated by a few cm and driven by a power generator. These discharges have been explored extensively over the past few decades. The power transfer mechanism, which is commonly referred to as 'electron heating' or 'electron power absorption' in the literature \citep{wilczek20:181101}, is still a topic rather poorly understood. Although the electron power absorption mechanism is a topic widely studied and discussed over the past decades, a fully consistent and general mathematical-physical explanation of the several physical mechanisms involved in the power transfer mechanism are still lacking. This is in particular true for the electronegative capacitively coupled discharge.

It is widely accepted that the electron heating can be divided into two components: the ohmic heating (collisional) and the stochastic heating (collisionless) while several operating modes have been identified in the capacitively coupled discharge including the stochastic electron heating due to the sheath motion ($\alpha$-mode) \citep{lieberman98:955}, secondary electron emission due to ion and neutral bombardment of the electrodes \citep{godyak86:112}, the drift ambipolar DA-mode \citep{schulze11:275001}, non linear electron resonance heating (NERH) \citep{mussenbrock06:151503, czarnetzki06:123503, mussenbrock08:085004, lieberman08:063505, donko09:131501}, electron bounce resonance effect \citep{park07:085003, liu11:055002} and the generation of series resonance oscillations \citep{mussenbrock06:151503, mussenbrock08:085004}.
In a strongly electronegative discharge the electrical conductivity tends to be low and, due to large ion inertia high electric field is induced within the plasma bulk (electronegative core). Furthermore, ambipolar fields appear near the sheath edges.


The particle-in-cell (PIC) method, when combined with Monte Carlo (MC) treatment of collision processes, is the most frequently used numerical approach to investigate the properties and the operating modes of low pressure capacitively-coupled-discharges. The combination of the particle-in-cell (PIC) method and Monte Carlo collision (MCC) treatment of collision processes is commonly referred to as the PIC/MCC method. 
 The PIC/MCC method is a self-consistent kinetic approach that has become a predominant numerical approach to investigate the properties of the low pressure capacitively-coupled discharge. 

The one-dimensional-object-oriented plasma device one ({\tt oopd1}) code allows having the simulated particles of different weights so that in principle both charged and neutral particles can be tracked during the simulation.  Earlier we benchmarked the basic reaction set for the oxygen discharge in the {\tt oopd1} code to the {\tt xpdpl} code \citep{gudmundsson13:035011}.
In recent years the oxygen reaction set in the {\tt oopd1}
code  has been improved significantly \citep{gudmundsson13:035011, gudmundsson15:035016, hannesdottir16:055002}.
The {\tt oopd1}
code has been applied to explore the electron power absorption in the capacitively coupled oxygen discharge while varying the various external parameters and operating conditions  such as discharge pressure
\citep{gudmundsson15:153302, hannesdottir16:055002, gudmundsson17:120001}, driving voltage amplitude \citep{gudmundsson17:193302}, driving frequency \citep{gudmundsson18:025009}, the secondary electron emission \citep{hannesdottir16:055002,proto18:65}, the surface quenching of the metastable states \citep{proto18:074002} and the electrode gap distance \citep{gudmundsson19:045012}.

During the past decades several attempts to describe correctly the behaviour of the electron heating using the Boltzmann equation have been made. Surendra and Dalvie \cite{surendra93:3914} were the first to set up a mathematical model to describe the electron power absorbed using the Boltzmann equation for both electrons and ions using the PIC results as input. 
In the years that followed several authors used the formulation set by \citet{surendra93:3914} to develop similar models inspired by the their results \citep{turner95:1312, gozadinos01:135004, lafleur14:035010, liu18:025006, brinkmann16:014001, grapperhaus97:569, kaganovich96:3818}. Among these Brinkmann \cite{brinkmann16:014001} derived a unified description of electron power absorption in capacitively coupled discharges using a mathematical formulation where the electron density profile has been approximated by a smooth step function, finding that the total time averaged electron power absorption is the sum of four terms, each one corresponding to one of the heating mechanism knows from separate previous theories, i.e.~NERH, stochastic heating (hard wall model), ambipolar/pressure heating and ohmic heating. Brinkmann also demonstrated that a time dependent temperature is necessary to obtain a non zero time averaged electron power absorption. More recently, Schulze et al used a simplified moment analysis of the Boltzmann equation (the Boltzmann term analysis) where the electron temperature gradient was both neglected and considered \citep{schulze08:105214, schulze18:055010} in order to describe both the electric field and the electron power absorbed in an electropositive low pressure capacitively coupled argon discharge. They found that the time averaged ambipolar electron power absorption completely vanishes for a temporally independent electron temperature. This approach has recently been applied to explore the electron power absorption mechanisms in a capacitively coupled oxygen discharge by \citet{vass20:025019}. Using the Boltzmann term analysis, they found that the ohmic contribution to the electron power absorption is small at different time steps at low pressure while it becomes important  at higher pressures. Finally, they observed that at low pressure the space-time averaged electron power absorbed was entirely given by the ohmic term and that the pressure term contribution increases as the pressure is increased.
Here we use the Boltzmann term analysis to investigate the origins of both the electric field and the power absorbed by the electrons at different time steps within both the plasma bulk and the sheath collapsed region and the related time averaged quantities within the plasma bulk in a capacitively coupled oxygen discharge at 10 and 100 mTorr for 45 mm of gap distance. The electron power absorption in the capacitively coupled oxygen discharge is distinctly different when the discharge is operated at 100 mTorr than when it is operated at 10 mTorr \citep{gudmundsson15:153302, hannesdottir16:055002, gudmundsson19:045012}. 
When operating at 100 mTorr with a gap size of 45 mm the discharge is operated in $\alpha$-mode and stochastic electron heating dominates while at 10 mTorr the discharge is more electronegative and the DA-mode dominates.  The main task of the current work is to perform a Boltzmann term analysis of a capacitively coupled oxygen discharge, in order to shed light on the the underlying physical mechanism behind the electric field and the electron power absorbed, to gain understanding of the electron power absorption in an  capacitively coupled oxygen discharge operated at different pressures (10 and 100 mTorr), that represent hybrid DA-$\alpha$ mode and pure $\alpha$ mode, respectively. We will follow the framework of the Boltzmann term analysis given in the recent work of \citet{schulze18:055010} with some modifications, since the physical conditions and gas considered are different. The current work is structured as follows. In Section \ref{Thesim} we give a brief overview of the simulation set up. In Section \ref{simulation} we show the spatio-temporal profiles of both the total charge density and the quasineutrality deviation along with both the total charge density and the density profiles for all the species involved at different time steps and time averaged. Section \ref{MDAR} discusses the model. In subsection \ref{SFM} a simple fluid model based on the \citet{schulze18:055010} work is employed to explore the behaviour of both the electric field and the electron power absorbed at both 100 and 10 mTorr. The results from both the simulations and the calculations at both 10 and 100 mTorr are discussed and compared in Section \ref{RaD}. Finally, Section \ref{Conlc} summarizes our findings.

\section{The Simulation}\label{Thesim}
The one-dimensional (1d-3v) object-oriented particle-in-cell Monte Carlo collision (PIC/MCC) code {\tt oopd1} \citep{gudmundsson13:035011} is applied to a  capacitively coupled oxygen discharge. In 1d-3v PIC codes the model system has one spatial dimension and three velocity components. The {\tt oopd1} code,  like the well known {\tt xpdp1} code, is a general plasma device simulation tool capable of simulating various types of devices, where the plasma is the main actor, such as particle beams, electrical breakdown,  particle accelerators as well as processing discharges \citep{gudmundsson13:035011}. The oxygen reaction set included in the {\tt oopd1} code is rather extensive and nine different species are considered: electrons,  the ground state  neutrals O($^3$P) and O$_2(\mathrm{X}^3\Sigma_{\rm g}^-)$, the negative ions O$^-$, the positive ions O$^+$ and O$_2^+$, and the metastables 
O($^1$D), O$_2(\mathrm{a}^1\Delta_{\rm g})$ and O$_2$(b$^1\Sigma_{\rm g}^+)$. The basic reaction set included O$_2(\mathrm{X}^3\Sigma_{\rm g}^-)$, O$_2^+$ and O$^-$. In our earlier work we added oxygen atoms in the ground state O($^3$P) and ions of the oxygen atom  O$^+$ to the {\tt oopd1} code \citep{gudmundsson13:035011}. In a later work the singlet metastable molecule O$_2(\mathrm{a}^1\Delta_{\rm g})$ and the metastable oxygen atom O($^1$D) were added \citep{gudmundsson15:035016}, as well as the singlet metastable molecule O$_2$(b$^1\Sigma_{\rm g}^+$) \citep{hannesdottir16:055002}. The full oxygen reaction set together with the cross sections used have been discussed in our earlier works and will not be repeated here \citep{gudmundsson13:035011,hannesdottir16:055002,gudmundsson15:035016,gudmundsson19:045012}. We assume a geometrically symmetric capacitively coupled discharge where one of the electrodes is driven by an rf voltage at a single frequency
\begin{align}
V(t)=V_0 \sin \left( 2 \pi f t \right)
\end{align}
while the other electrode is grounded. Here, $V_0$ is the voltage amplitude, $f$ the driving frequency and $t$ the time. For this current study we assume the discharge to be operated at the pressure of 10 mTorr and 100 mTorr with voltage amplitude $V_0=$400 V with an electrode separation of 4.5 cm. A capacitor of 0.1 $\mu$F is connected in series with the voltage source. The electrode diameter and the driving frequency are assumed to be 10.25 cm and 13.56 MHz respectively. These are the same parameters as assumed in our previous work \cite{gudmundsson19:045012}. The time step $\Delta t$ and the grid spacing $\Delta x$ are set to resolve the electron plasma frequency and the electron Debye length of the low energy electrons respectively, according to $\omega_{\rm pe} \Delta t < 0.2$, where $\omega_{\rm pe}$ is the electron plasma frequency and the simulation grid consists of 1000 equal cells. The electron time step is $3.68 \times 10^{-11}$ s.  The simulation was run for $5.5 \times 10^6$ time steps, which corresponds to 2750 rf cycles as it takes roughly 1700 rf cycles to reach equilibrium for all particles. Time averaged plasma parameters shown, such as the densities, the electron power absorption, and the effective electron temperature, are averaged over 1000 rf cycles. All particle interactions are treated by the Monte Carlo method with a null-collision scheme \citep{birdsall91:65}. For the heavy particles, we apply sub-cycling, where the heavy particles are advanced every 16 electron time steps \citep{kawamura00:413} and an initial parabolic density profile has been assumed \citep{kawamura00:413}.

The kinetics of the charged particles (electrons, O$_2^+$ ions, O$^+$ ions and O$^-$ ions) was followed for all energies. Since the neutral gas density is much higher than the densities of charged species, the neutral species at thermal energies (below a certain cut-off energy) are treated as a background with fixed density and temperature and maintained uniformly in space.  The main challenge when PIC/MCC simulations are applied to simulate molecular gases  has to due with the timescale difference between the processes of dissociation and the processes involving charged particles. Therefore, a global model~\citep{thorsteinsson10:055008} is used beforehand to determine the partial pressure of the various neutrals created in the discharge as discussed in Proto and Gudmundsson \citep{proto18:074002}, i.e.~the ground state neutral atoms O($^3$P) and the metastables O($^1$D), O$_2(\mathrm{a}^1\Delta_g)$ and O$_2$(b$^1\Sigma_g^+)$ under certain control parameters including the discharge pressure, the absorbed power and the gap separation between the two electrodes, etc. The absorbed power determined by the PIC/MCC simulation is used as an input parameter in the global model calculations, iteratively. The partial pressure of the atoms and metastable species obtained from the global model calculation is then used as the partial pressure of these species in the neutral background gas in the simulation. Note that, a global model is mainly developed to model a high density low pressure discharges such as inductively coupled discharges, rather than capacitively coupled discharges and the proportion of the power absorbed by the  electrons in the former is much larger than in the latter. Therefore, the global model may overestimate  the atom and metastable density within the discharge, especially when operating at low pressure. The fractional densities for the neutrals O$_2(\mathrm{X}^3 \Sigma_{\rm g}^-)$, O$_2$($\mathrm{a}^1\Delta_{\rm g}$), O$_2$($\mathrm{b}^1\Sigma_{\rm g}$), O($^3$P), O($^1$D), estimated using the global model calculations, are listed in Table \ref{partialpressure}. These values have been used as input for the PIC/MCC simulation as the partial pressures of the neutral background gas.  These neutral background species are assumed to have a Maxwellian velocity distribution at the gas temperature (here $T_n=$26 meV). The kinetics of the neutrals are followed when their energy exceeds a preset energy threshold value. The energy threshold values used here for the various neutral species are listed in Table \ref{simparr}. The thresholds were chosen in order to keep the number of simulated particles within a suitable range, typically $10^4-10^5$ particles. Particles with energy below this threshold energy are assumed to belong to the neutral background.

 Note that the background neutrals are assumed to be uniform within the discharge.  However, we are aware that the electrode surfaces have a significant influence on the neutral density profiles. The density profiles for fast neutrals indicate that the oxygen atom density decreases and the molecular metastable density increases in the electrode vicinity \citep{hannesdottir16:055002}. 
As an oxygen atom O($^3$P) hits the electrode, it is assumed that half of the atoms are reflected as O($^3$P) at room temperature and the other half recombines to from the ground state oxygen molecule O$_2(\mathrm{X}^3 \Sigma_{\rm g}^-)$ at room temperature. Similarly, as a metastable oxygen atom O($^1$D) hits the electrode, half of the atoms are quenched to form O($^3$P) and the other half, is assumed to recombine to form the ground state oxygen molecule O$_2(\mathrm{X}^3 \Sigma_{\rm g}^-)$ at room temperature. The surface quenching coefficients for the singlet metastable molecules on the electrode surfaces are assumed to have a value of $\gamma_{\rm wqa}=0.007$ and $\gamma_{\rm wqb}=0.1$ for O$_2$($\mathrm{a}^1\Delta_{\rm g}$) and O$_2$($\mathrm{b}^1\Sigma_{\rm g}$), respectively. 
The influence of the surface quenching coefficients of the singlet metastable molecule on the electron heating mechanism has been explored in detail in an earlier work \citep{proto18:074002}, where it has been demonstrated that the influence of $\gamma_{\rm wqa}$ on the overall discharge properties can be rather significant. The surface quenching and recombination coefficients used in this current work are listed in Table \ref{simparr}.  Note that the  oxygen reaction set used in this current study is signficantly more extensive than the one used by \citet{vass20:025019} where the only metastable state included is the singlet metastable molecule O$_2(^1\Delta_{\rm g})$. To estimate the density of the singlet metastable molecule O$_2(^1\Delta_{\rm g})$ they assume a homogeneous  spatial density where a balance between electron  impact excitation and  quenching at the electrodes \citep{derzsi16:015004,wang19:055007}.

Secondary electron emission and electron reflection has been incorporated into the discharge model \citep{hannesdottir16:055002}. The energy dependent secondary electron emission yield for a dirty surface has been employed \citep{hannesdottir16:055002} and the electrons are assumed to be elastically reflected from the electrodes independently of their energy and their angle of incidence with a probability of 0.2 \citep{proto18:65}. The secondary electron emission due to the electron impact on the electrodes has been neglected here as in all our previous works on the oxygen discharge.
\begin{table}[!ht]
\renewcommand\arraystretch{1.5}
		\caption{\label{partialpressure} The relative partial pressures of the thermal neutrals at 10 and 100 mTorr  calculated by a global (volume averaged) model for 1.8 W power. }
		\begin{tabular}{ c  c  c  c  c }
		\hline \hline
             &	O$_2$$(\mathrm{X}^3 \Sigma_{\rm g}^-)$	&	 O$_2$(a$^1 \Delta_{\rm g})$ & O$_2$(b$^1 \Sigma_{\rm g})$ & O$(^3$P) \\ \hline
10 mTorr   	&  $0.9801$	& $0.0148$	& $0.0016$ & $0.0008$  \\	\hline
100 mTorr   	&  $0.9877$	& $0.0017$	& $0.0009$ & $0.0095$ \\
\hline \hline
		\end{tabular}
\end{table}
\begin{table*}
\caption{The parameters of the simulation, the particle weight, the  energy threshold above 
which kinetics of the neutral particles are followed and the wall 
recombination and quenching coefficients for the neutral species on the electrode surfaces.
}
 \label{simparr}
 \begin{tabular}{llll}
\hline \hline
Species & particle weight  & energy   threshold & coefficient \\
        & (range) &   [meV]   & recomb./quenching \\ 
\hline
 O$_2(\mathrm{X}^3 \Sigma_{\rm g}^-)$     &  $5 \times 10^7$  & 500    &            \\    
 O$_2$($\mathrm{a}^1\Delta_{\rm g}$)      &  $5 \times 10^6$  & 100 & 0.007    \\    
 O$_2$($\mathrm{b}^1\Sigma_{\rm g}$)      &  $5 \times 10^6$  & 100 & 0.1    \\    
  O($^3$P)                          &  $5 \times 10^7$  & 500  & 0.5       \\    
  O($^1$D)                          &  $5 \times 10^7$  & 50   & 0.5 recomb/0.5 quenching         \\ 
      O$_2^+$                       &   $10^7 - 10^8$          & -        \\ 
      O$^+$                         &   $10^6 - 10^7$          & -       \\ 
      O$^-$                         &   $5 \times 10^7 - 10^8$ & -      \\ 
      e                            &  $10^7 - 10^8$  & -      \\ 
\hline \hline         
 \end{tabular} 
\end{table*} 
\section{Results from the Simulation}\label{simulation}
Through recent PIC/MCC simulations of a capacitively coupled oxygen discharge it has been demonstrated that the singlet metastable molecular states have a significant influence on the electron power absorption mechanism \citep{gudmundsson15:035016,gudmundsson15:153302,hannesdottir16:055002,gudmundsson17:120001} 
as well as the ion energy  distribution \citep{hannesdottir17:175201}. At low (high) pressure and high (low) electronegativity, i.e.~10 mTorr (50 -- 500 mTorr), the electron power absorption is mainly located within the plasma bulk (the sheath regions) \citep{gudmundsson15:153302, hannesdottir16:055002}. Furthermore, when operating at low pressure, the time averaged electron power absorption within the discharge is due to a hybrid drift-ambipolar mode,  (DA-mode) and $\alpha$-mode, and while operating at higher pressures, the electron power absorption is due to stochastic heating and the discharge is operated in a pure  $\alpha$-mode \citep{gudmundsson17:193302, gudmundsson18:025009}.
It has also been demonstrated that detachment by singlet molecular metastable states is the process that has the most influence on the electron power absorption process in the higher pressure regime,  while it has almost negligible influence at lower pressures \citep{hannesdottir16:055002, gudmundsson15:153302, gudmundsson17:120001}.  All the quantities returned by the simulations and involved in the calculations for both the electric field and the electron power absorption in the following sections are arrays extended along the $x-$axis, i.e.~the discharge gap length. In particular, every single component of the electron temperature has been calculated as $T_{{\rm e}, ii}=\frac{2}{e} {\cal E}_{{\rm e}, ii} - \frac{m_{\rm e}}{e} u_{{\rm e}, i}^2$ where ${\cal E}_{{\rm e}, ii}$ and $u_{{\rm e}, ii}^2$, with $i=x, y, z$, are the mean electron energy density and the electron mean velocity respectively. Since ${\cal E}_{{\rm e}, i}=\frac{m_{\rm e}}{2} \langle v_{{\rm e}, i}^2 \rangle$ by definition, the expression for the electron temperature given above is the same as the one shown by \citet{wilczek20:181101}, when the particle mean velocity is not negligible.

\begin{figure}[!ht]
\includegraphics[width=8 cm]{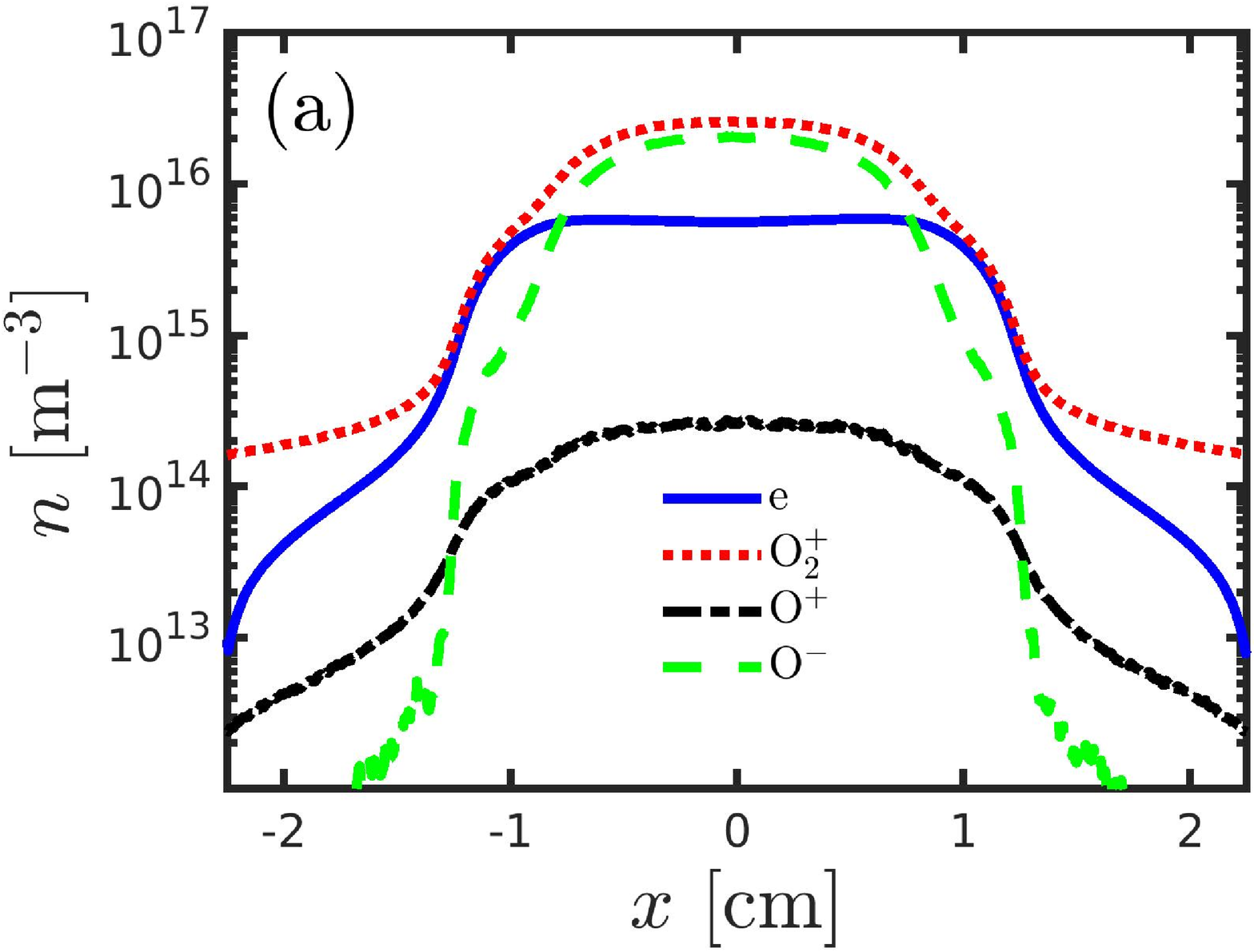}   
\includegraphics[width=8 cm]{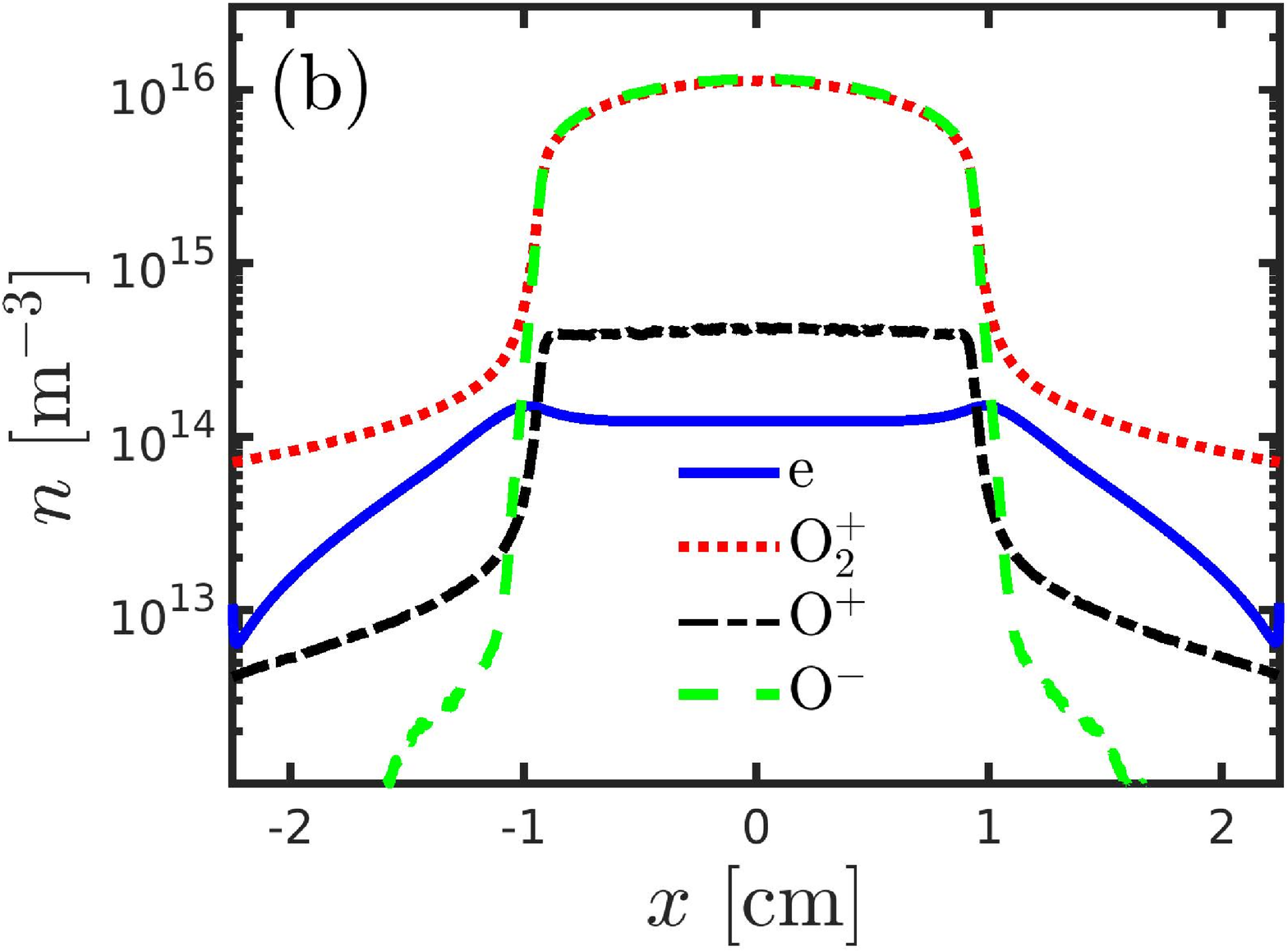} 
\caption{\label{figZ}The density profiles for charged particles at (a) 100 mTorr and at (b) 10 mTorr  in a parallel plate capacitively coupled discharge  with a gap
separation of 45 mm  driven by a 400 V voltage source at 13.56 MHz.}
\end{figure}

Figure \ref{figZ} shows the density profiles for  O$_2^+$ ions, O$^+$ ions, O$^-$ ions and electrons at 100 and 10 mTorr. At 100 mTorr the center electronegativity is 3.55 and at 10 mTorr it is 93.64.
The electronegative discharge consists of an electronegative core connected to electropositive edge plasma regions \citep{lichtenberg94:2339, lichtenberg97:437}. At 100 mTorr (Figure \ref{figZ} (a)) both O$^-$ ion and O$_2^+$ ion density profiles have a similar shape within the bulk region while the O$^-$ ion density decreases more steeply than the O$_2^+$ ion density profile approaching the sheath edges. We also see that the electron density profile is somewhat lower than both the O$^-$ ion and the O$_2^+$ ion density profiles within the bulk region while it decreases sharply within both the sheath regions. At 10 mTorr (Figure \ref{figZ} (b)) the situation is different. First of all we see that O$^-$ and the O$_2^+$ ion density profiles perfectly overlap within the bulk region and that the O$^-$ ion density profile decreases more steeply than the O$_2^+$ density profile beyond the sheath edges. The time averaged value of both the O$^-$ and the O$_2^+$ ion density profiles within the bulk region is slightly lower than in the 100 mTorr case. The electron density in the bulk plasma is roughly two orders of magnitude lower than the O$_2^+$ and O$^-$ densities. Moreover, we observe that the O$^+$ density is higher than the electron density within the bulk region,  contrary to the 100 mTorr case. We also observe that the O$^+$ density profile is more flattened within the bulk region and that it decreases more steeply than for the 100 mTorr case. The electron density profile is flat and constant within the bulk region and it has equal absolute maxima on both the sheath edges.
\begin{figure}[!ht]
\begin{center}
\includegraphics[width=8 cm]{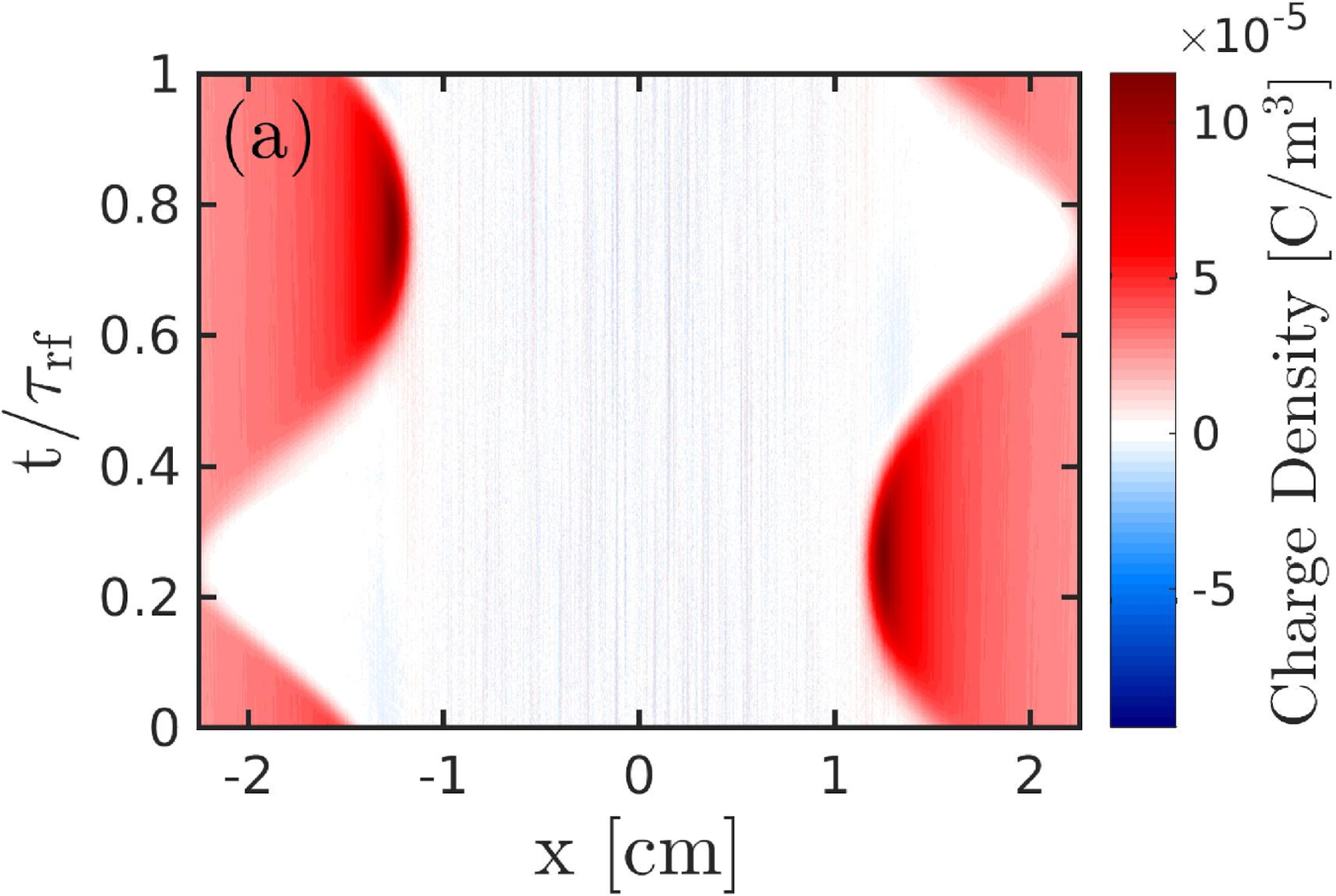}   
\includegraphics[width=8 cm]{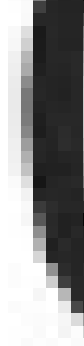}   
\end{center}
\caption{\label{figA}The spatio-temporal behaviour of the (a) total charge density and of the (b) quasineutrality deviation, defined by Eqs.~\eqref{results1} and \eqref{results2}, respectively,  over the full gap length for a parallel plate capacitively coupled oxygen discharge at 100 mTorr for 45 mm of gap separation driven by a 400 V voltage source at driving frequency of 13.56 MHz.}
\end{figure}
Figure \ref{figA} (a) shows the spatio-temporal behaviour of the total charge density at 100 mTorr over the full gap length for a full period defined as follows
\begin{align}\label{results1}
\text{Total Charge Density}=e \left( n_{\rm O_2^+} + n_{\rm O^+} - n_{\rm O^-} - n_{\rm e} \right)
\end{align}
Firstly, we observe a net zero charge density within the bulk region and the fully collapsed sheath regions. Secondly, the right (left) sheath region is positively charged  and reaches its maximum extension at $t / \tau_{\rm rf}=0.25$ ($t / \tau_{\rm rf}=0.75$). Moreover, the right (left) sheath positive net charge has a peak on the bulk side of the right (left) sheath edge and slowly decreases while approaching the right (left) electrode. Figure \ref{figA} (b) shows the spatio-temporal behaviour of the quasineutrality deviation at 100 mTorr over the full gap length defined as follows
\begin{align}\label{results2}
\text{Quasineutrality deviation}=\frac{\left( n_{\rm O_2^+} + n_{\rm O^+} - n_{\rm O^-} - n_{\rm e} \right)}{n_{\rm O_2^+} + n_{\rm O^+}} .
\end{align}
We observe that the quasineutrality deviation uniquely identifies the sheath region. Indeed, the quasineutrality deviation value is $1$ within the expanded sheaths while it is $0$ within the plasma bulk and the sheath collapse regions and it has an intermediate value on the bulk-sheath time varying interface.
\begin{figure}[!ht]
\begin{center}
\includegraphics[width=8 cm]{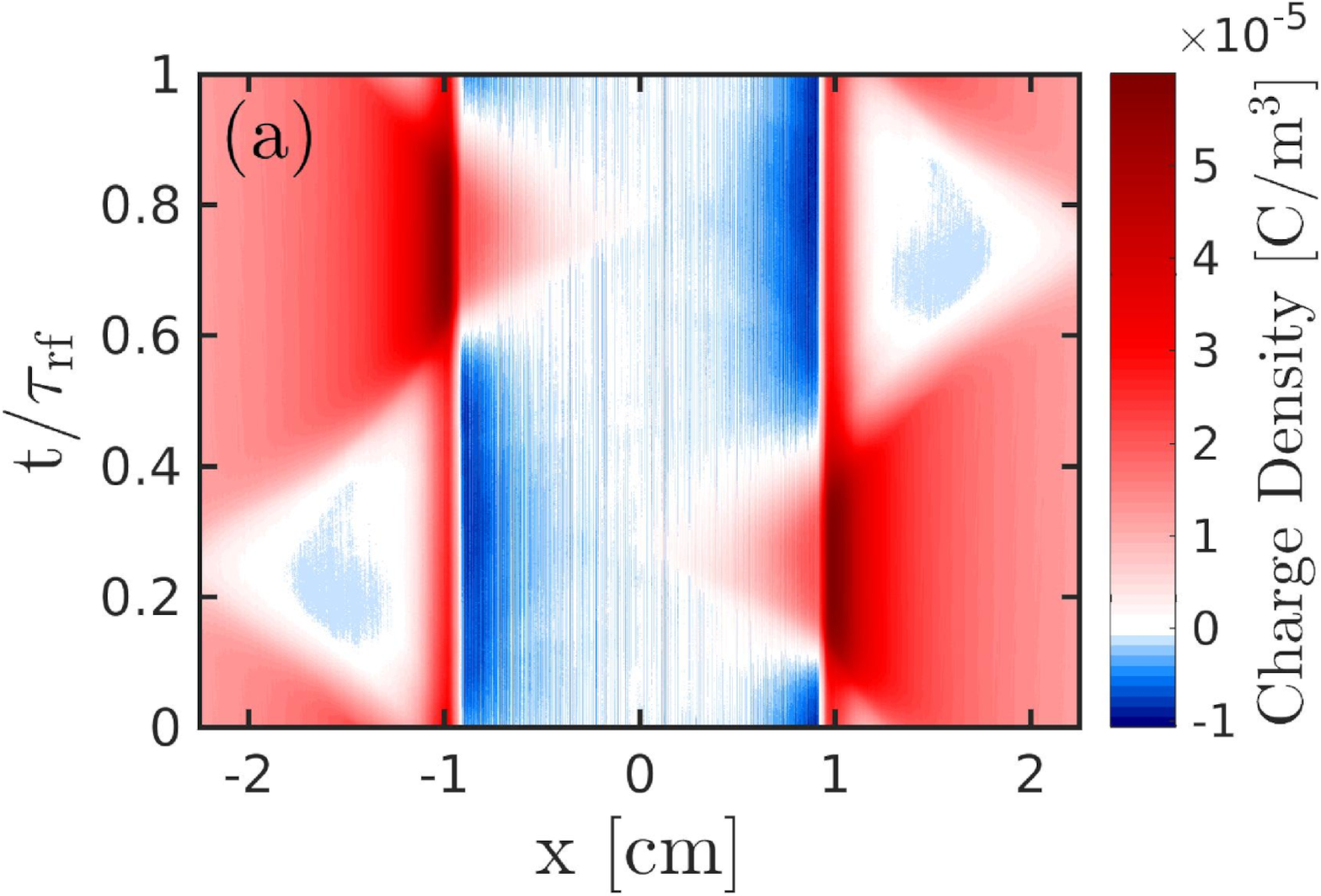}   
\includegraphics[width=8 cm]{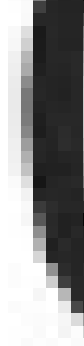}   
\end{center}
\caption{\label{figA2}The spatio-temporal behaviour of the (a) total charge density and the (b) quasineutrality deviation, defined by Eqs.~\eqref{results1} and \eqref{results2}, respectively,  over the full gap length for a parallel plate capacitively coupled oxygen discharge at 10 mTorr for 45 mm of gap separation driven by a 400 V voltage source at driving frequency of 13.56 MHz.}
\end{figure}
Figure \ref{figA2} (a) shows the spatio-temporal behaviour of the total charge density at 10 mTorr over the full gap length for a full period defined by Eq.~\eqref{results1}.  Firstly, we observe a net positive (negative) charged stripe that appears on the sheath side (on the bulk side) of both the sheath edges over the full rf-cycle (on both the collapsing sheath edges). The positive charged stripe density strongly increases on the sheath side of the expanded sheath edge. This positive charged stripe was absent in the 100 mTorr case as shown in Figure  \ref{figA} (a). Such a difference with respect to the 100 mTorr case is due to the fact that the electron mean free path is longer at low pressures, so that they leave the positive ions behind while crossing the sheath edge during the sheath collapse. For example, at $t / \tau_{\rm rf}=0.25$, on very short time scales, a net negative ambipolar field builds up (Figure \ref{figA2} (a)), which induces a recall force on the bulk electrons and a pushing force on the bulk positive ions toward the collapsing sheath edge. For this reason, an excess of positive charges on the immediate sheath side of both the collapsed sheath edges is observed. It's worth noting that, once crossed the sheath side of the collapsed sheath edge, the electrons are free to accelerate towards the left electrode due to the flux compensation effect. On the other hand, a positive peak in the ambipolar field is observed on the bulk side of the collapsed sheath edge at 10 mTorr (Figure \ref{figA2} (a)). Such a peak is absent at 100 mTorr (Figure \ref{figA} (a)). This can be explained considering that, at longer time scales, when the electrons are repelled from the sheath side of the collapsed sheath edge, the ambipolar field changes the sign and becomes positive on the bulk side of the collapsed sheath edge. Such a behaviour is responsible for the negative charge excess observed on the bulk side of the collapsed sheath edge. Moreover, under the action of the ambipolar force, the electrons are confined within the bulk and the collapsed sheath region. The same reasoning can be also applied at both $t / \tau_{\rm rf}=0.50$ and $t / \tau_{\rm rf}=0.75$. We also observe that the sheath regions reach a larger extension at both $t / \tau_{\rm rf}=0.25$ and $t / \tau_{\rm rf}=0.75$ with respect to the 100 mTorr case, while the net positive charge present on the sheath side of both the sheath edges has a lower value when the sheath approaches its maximum extension. The net charge in the fully expanded sheath regions at 10 mTorr is globally lower than in the 100 mTorr case. Figure \ref{figA2} (b) shows the spatio-temporal behaviour of the quasineutrality deviation at 10 mTorr over the full gap length defined as in Eq.~\eqref{results2}. As in the 100 mTorr case shown in Figure \ref{figA} (b),  the quasineutrality deviation uniquely identifies the sheath region. We observe a non-quasineutral stripe on both the sheath edges which is related to the positive charged stripe observed in Figure \ref{figA2} (a). At 100 mTorr the sheath has a smooth contour over the full rf cycle (Figure \ref{figA} (b)) while at 10 mTorr it ends abruptly on the non-quasineutral stripe on both the sheath edges.
\begin{figure}[!ht]
\begin{center}
\includegraphics[width=8 cm]{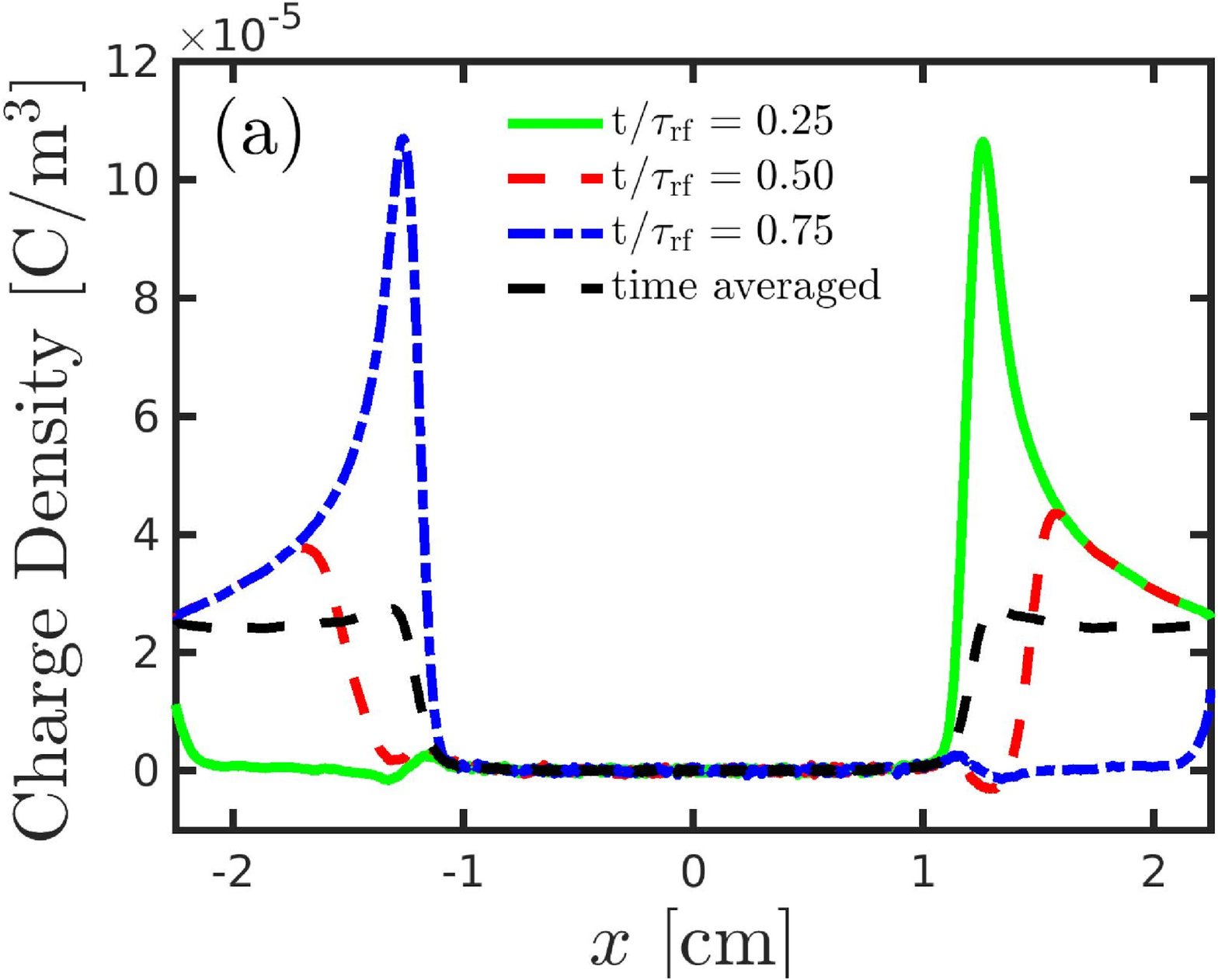} 
\includegraphics[width=8 cm]{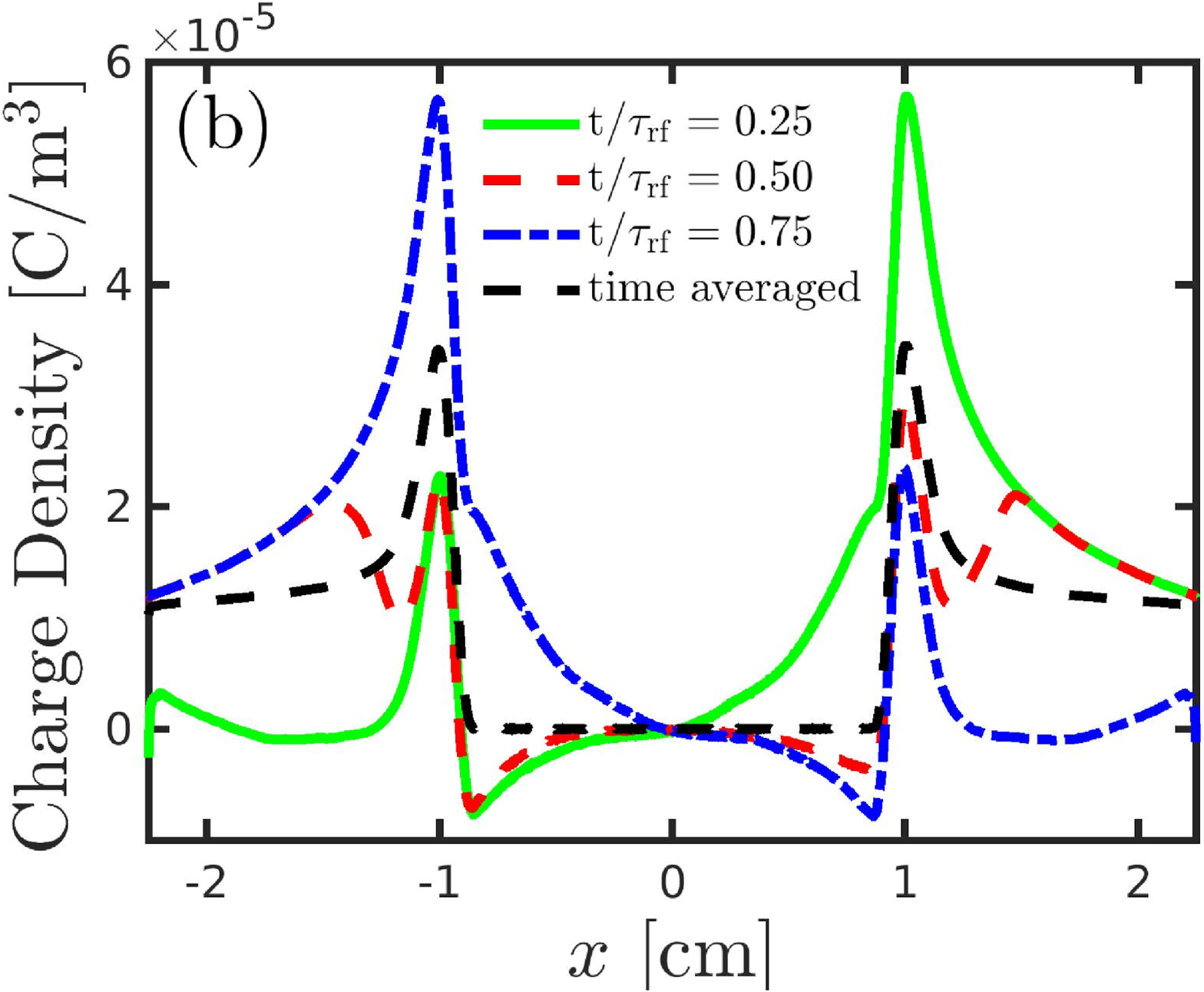} 
\end{center}
\caption{\label{figB}The total charge density at (a) 100 mTorr and at (b) 10 mTorr over the full gap length at $t / \tau_{\rm rf}=0.25$ (green line), $t / \tau_{\rm rf}=0.50$ (red dashed line), $t / \tau_{\rm rf}=0.75$ (blue dotted dashed line) and time averaged (black dashed line) for a parallel plate capacitively coupled oxygen discharge for 45 mm of gap separation driven by a 400 V voltage source at driving frequency of 13.56 MHz.}
\end{figure}
Figure \ref{figB} (a) shows the total charge density profile at 100 mTorr defined by Eq.~\eqref{results1} at different time steps and time averaged over all the full gap length. We observe that the total charge density profile is flat and very close to zero over the full bulk width for all the cases considered. At $t / \tau_{\rm rf}=0.25$ ($t / \tau_{\rm rf}=0.75$) the total charge density steeply increases while approaching the sheath edge of the right (left) bulk-sheath interface and sharply decreases toward the right (left) electrode. On the other hand, the total charge density is approximatively constant at $t / \tau_{\rm rf}=0.25$ ($t / \tau_{\rm rf}=0.75$) and zero within the fully collapsed sheath region and slowly increases toward the right (left) electrode. The charge density profile at  $t / \tau_{\rm rf}=0.25$ is a mirror image of the charged density profile at  $t / \tau_{\rm rf}=0.75$. At $t / \tau_{\rm rf}=0.50$ the total charge density slightly decreases once passed both the sheath edges and steeply increases on the sheath side of both the sheath edges. Then it slowly decreases while approaching both the electrodes overlapping with both the $t / \tau_{\rm rf}=0.25$ and the $t / \tau_{\rm rf}=0.75$ cases. In the time averaged case the total charge density steeply increases once passed both the sheath edges reaching a lower maximum with respect to all the other cases considered. We observe the time averaged case to be almost zero within the plasma bulk and to have an almost constant profile within the sheath regions while approaching both the electrodes.

Figure \ref{figB} (b) shows the total charge density profile at 10 mTorr defined by Eq.~\eqref{results2} at different time steps and time averaged over the full gap length. The total charge density is zero within the discharge center ($x=0$) for all the four cases considered even thought there is always some charge density within the plasma bulk (either positive or negative), and the time average is zero. Also, the total charge density profile at $t / \tau_{\rm rf}=0.25$ ($t / \tau_{\rm rf}=0.75$) shows an additional local maximum on the left (right) sheath edge with respect to the 100 mTorr case (Fig.~\ref{figB} (a)). The same applies to the $t / \tau_{\rm rf}=0.50$ and the time averaged case, with the presence of two almost equal local maxima on both the sheath edges, which are absent in the 100 mTorr case. At $t / \tau_{\rm rf}=0.25$ the total charge density profile sharply increases (decreases) while approaching the bulk side of the right (left) sheath edge and it has an absolute maximum (minimum) on the sheath side (bulk side) of the right (left) sheath edge. Then the total charge density profile sharply decreases (steeply increases) once passed the right (left) sheath edge reaching a positive value (a local maximum) toward the right electrode (on the sheath side of the left sheath edge). We also observe that the total charge density profile is approximatively constant within the left sheath region.
\section{Model description and Results}\label{MDAR}
During the past years several attempts have been made to describe correctly the behaviour of the electric field using the Boltzmann equation. Surendra and Dalvie \cite{surendra93:3914} used the first momentum Boltzmann equation to decompose the electric field into a sum of different terms, each one corresponding to different physical mechanisms. They were able to isolate all the single terms contributing to both the electric field and the electron power absorbed. Moreover, they found the electron pressure term to be important for the collisionless heating and that, for a constant electron temperature, the collisionless electron heating vanishes upon time average. Since then several attempts have been made to explain the behaviour of the electric field at different time steps and upon time average. In recent years Surendra's framework has been employed to explain the behaviour of the electric field within the bulk and in the sheath regions. 
In particular Schulze and coworkers have used the zeroth momentum Boltzmann equation (stationary continuity equation), with a stationary density profile \cite{schulze08:105214} and with a temporally dependent density profile together with a non zero ionization rate \cite{schulze18:055010}, combined with the first momentum Boltzmann equation, with and without the change in the momentum term \citep{schulze08:105214, schulze18:055010} to derive a space- and time-resolved expression for the different electric field terms involved. The Surendra-Dalvie framework has improved our knowledge of the physical mechanisms behind the origins of the electric field within the bulk region and within the collapsed sheath region but has not given a general consensus on the origin of the electric field within the expanded sheath region.

The DA-mode is associated with the creation of electric field within the plasma bulk.
The electric field within a plasma discharge is built up by several different phenomena, depending on the gas considered. The electronegative discharges present a bigger number of phenomena than the electropositive discharges, and the situation is much more complicated. For both electropositive and electronegative discharges sheaths form near the electrodes,
a positive net charge within the sheath region builds up, leading to a potential profile that is positive within the bulk region and falls to zero near both electrodes \citep{lieberman05}. 
However, a strong electric field within the bulk region has been observed, both experimentally and by simulations in electronegative discharges.
The high value of the electric field has been related to the low dc conductivity within the bulk as discussed by \citet{schulze11:275001}. Furthermore, strong peaks in the electric field at the sheath edges have been observed \citep{georgieva03:2369}. The observed peaks have been related to the corresponding local maxima of the electron density at the sheath edges which are caused by the ambipolar field built up by a net charge separation between the positive charges as they are accelerated towards the electrode,  and the electrons, together with the negative ions, confined within the bulk region. This is completely different from the situation observed in the electropositive discharges, where the ambipolar field accelerates the electrons toward the discharge center \citep{schulze11:275001}.

\subsection{Simple Fluid Model}\label{SFM}

When operated at 100 mTorr (10 mTorr) the electronegativity is low (high) and the discharge operates in pure $\alpha$-mode (hybrid DA-mode and $\alpha$-mode). Irrespectively of the different discharge modes and conditions, a simple fluid model for an electronegative discharge is sufficient to describe the physics of a such system. In this subsection the simple fluid model applied to a discharge operated at both 100 mTorr 10 mTorr is discussed.  The model describes the behaviour of the electric field and of the electron power absorption within the bulk region.
This model is based on the approach used by \citet{schulze18:055010}, with the only difference that here both the ionization rate and the change of momentum terms are assumed to be negligible and are set equal to zero. The model is valid within the bulk region and in the collapsed sheath regions only, since the electron density in the expanded sheath region is very small. Moreover, the quasineutrality condition has not been imposed and the ideal gas law has been employed in the first momentum Boltzmann equation.  At 100 mTorr the pressure tensor (the temperature) is taken as (not) isotropic. Setting $p_{\rm e} \equiv p_{\rm e, xx} = p_{\rm e, yy} = p_{\rm e, zz}$, one finds Tr$(p_{{\rm e}, ij})=3 p_{\rm e, xx}=3 p_{\rm e} = 3 e n_{\rm e} T_{\rm e}=e n_{\rm e} \left(T_{\rm e, xx} + T_{\rm e, yy} + T_{\rm e, zz} \right)$, so that $T_{\rm e} \equiv \left( T_{\rm e, xx} + T_{\rm e, yy} + T_{\rm e, zz} \right) / 3$, i.e.~the electron temperature is direction averaged. Accordingly to the current set up and in order to make the physical system consistent, the ideal gas law has to be seen as an approximation. On the other hand at 10 mTorr neither the pressure tensor, nor the temperature are isotropic. Since $p_{\rm e, xx} \neq p_{\rm e, yy} \neq p_{\rm e, zz}$ and $T_{\rm e, xx} \neq T_{\rm e, yy} \neq T_{\rm e, zz}$, we are left with $p_{\rm e} \equiv p_{\rm e, xx}=e n_{\rm e, xx}T_{\rm e, xx}, p_{\rm e, yy}=e n_{\rm e, yy}T_{\rm e, yy}$, and $p_{\rm e, zz}=e n_{\rm e, zz}T_{\rm e, zz}$. The zeroth and the first momentum Boltzmann equation for electrons in a plasma discharge in the absence of magnetic field are the continuity equation
\begin{align}
\label{equation1}
 \frac{\partial n_{\rm e}}{\partial t} + \frac{\partial}{\partial x} \left( u_{\rm e} n_{\rm e} \right) = G - L
\end{align}
 where $G$ and $L$ are the reaction rates involving the creation and the destruction of electrons, respectively, and the momentum balance equation
\begin{align}
\label{equation2}
&\frac{\partial}{\partial t} \left[ m_e n_e u_e \right] + \frac{\partial}{\partial x} \left[ m_e n_e u_e^2 \right] + \frac{\partial}{\partial x} \left[ e n_e T_e \right] \nonumber \\  &+ e n_e E +  \Pi_{\rm c}=0,
\end{align}
respectively.  According to this set up, there is no need for keeping the continuity equation (Eq. \eqref{equation1}). Now, according to \citet{lieberman05}, the momentum change term $\Pi_{\rm c}$ can be approximated by a Krook collisional operator as follows
\begin{align}\label{equationextra1}
 \Pi_{\rm c}=\sum_{\beta} m_{\rm e} n_{\rm e} \nu_{\rm e \beta} \left( u_{\rm e} - u_{\rm \beta} \right) - m_{\rm e} \left( u_{\rm e} - u_{\rm G} \right) G + m_{\rm e} \left( u_{\rm e} - u_{\rm L} \right) L
\end{align}
where the summation is over all species,$u_{\rm e}$ and $u_{\rm \beta}$ are the mean velocities of the electrons and the species $\beta$, respectively, and $\nu_{\rm e \beta}$ is the momentum transfer frequency for collisions between electrons and species $\beta$. Now, neglecting the reactions involving the creation and destructions of particles (e.g., ionization, recombination) and considering only the O$_2$ neutral species, with a negligible velocity compared to the electrons, the momentum change term becomes
\begin{align}\label{equationextra2}
 \Pi_{\rm c}= m_{\rm e} \nu_{\rm e} n_{\rm e} u_{\rm e}
\end{align}
along with the continuity equation
\begin{align}\label{equationextra3}
 \frac{\partial n_{\rm e}}{\partial t} + \frac{\partial}{\partial x} \left( u_{\rm e} n_{\rm e} \right) = 0
\end{align}
Solving Eq.~\eqref{equationextra3} with respect to the velocity gradient one finds
\begin{align}
\label{equation3}
\frac{\partial u_e}{\partial x}= - \frac{u_e}{n_e} \frac{\partial n_e}{\partial x} - \frac{1}{n_e} \frac{\partial n_e}{\partial t}
\end{align}
Combining Eqs.~\eqref{equation3},  \eqref{equationextra2} and \eqref{equation2} together with the ideal gas law one finds an expression for the electric field
\begin{align}\label{equation4}
E=&- \underbrace{\frac{m_{\rm e}}{e} \frac{\partial u_{\rm e}}{\partial t}}_{\rm I} + \underbrace{\frac{m_{\rm e}}{e} \frac{u_{\rm e}^2}{n_{\rm e}} \frac{\partial n_{\rm e}}{\partial x}}_{\rm II} + \underbrace{\frac{m_{\rm e}}{e} \frac{u_{\rm e}}{n_{\rm e}} \frac{\partial n_{\rm e}}{\partial t}}_{\rm III} - \underbrace{\frac{T_{\rm e}}{n_{\rm e}} \frac{\partial n_{\rm e}}{\partial x}}_{\rm IV} \nonumber \\ &- \underbrace{\frac{\partial T_{\rm e}}{\partial x}}_{\rm V} - \underbrace{\frac{m_{\rm e} u_{\rm e} \nu_c}{e}}_{\rm VI}
\end{align}
Each electric field term in Eq.~\eqref{equation4} has its own origin. The first and the third term (I and III) are electron inertia terms due to the temporal variation in the electron velocity and density, respectively. The second term (II) corresponds to an electric field due to the normalized electron density gradient.
The fourth (IV) term corresponds to diffusion (ambipolar field) \cite{schulze11:275001, schulze08:105214}. The fifth term (V) corresponds to the electron temperature gradient. Therefore, terms IV and V represent electron heating due to pressure effects which is a collisionless mechanism \citep{turner95:1312}. The sixth term (VI) is due to electron collisions with atoms and molecules (drift field). Equation \eqref{equation4} has been applied to a given set of input parameters. The input parameters are the electron density and the electron temperature from the simulation. The collision term (Term VI) was taken from the reaction rate given by the simulation for an electron neutral elastic collision. The electron collision frequency values at 100 and 10 mTorr are $\nu_c=8.16 \times 10^7$ s$^{-1}$ and $\nu_c=5.06 \times 10^7$ s$^{-1}$ within the discharge center, respectively.  Multiplying the electric field coming from Eq.~\eqref{equation4} times the electron current density $J_e=-e n_{\rm e} u_{\rm e}$ it is possible to find the electron absorbed power as follows
\begin{align}\label{equation5}
J_{\rm e} \cdot E=&\underbrace{m_{\rm e} u_{\rm e} n_{\rm e} \frac{\partial u_{\rm e}}{\partial t}}_{\rm I} - \underbrace{m_{\rm e} u_{\rm e}^3 \frac{\partial n_{\rm e}}{\partial x}}_{\rm II} - \underbrace{m_{\rm e} u_{\rm e}^2 \frac{\partial n_{\rm e}}{\partial t}}_{\rm III} \nonumber \\ &+ \underbrace{e u_{\rm e} T_{\rm e} \frac{\partial n_{\rm e}}{\partial x}}_{\rm IV} + \underbrace{e n_{\rm e} u_{\rm e} \frac{\partial T_{\rm e}}{\partial x}}_{\rm V} + \underbrace{m_{\rm e} n_{\rm e} \nu_{\rm c} u_{\rm e}^2}_{\rm VI}
\end{align}
Each electron power absorption term that constitutes Eq.~\eqref{equation5} has its own origin which is strictly related to the electric field given by Eq.~\eqref{equation4}. The first and the third term (I and III) are electron inertia power absorption terms. The second term (II) corresponds to the power absorption term related to the electron density gradient.
The fourth (IV) term is related to the ambipolar field orignating from the electron density gradient \cite{schulze11:275001, schulze08:105214}. The fifth term (V) is related to the electron temperature gradient term for the electric field (Eq.~\eqref{equation4}). The fourth and fifth terms are usually know in the literature as pressure heating terms, respectively \cite{schulze18:055010, turner95:1312, surendra93:3914}. The sixth term (VI) is related to the collisions and represents ohmic heating. It's worth noting that the electron power absorption formula shown in Eq.~\eqref{equation5} can be split as follows \cite{surendra93:3914}
\begin{align}\label{split1}
(J_{\rm e} \cdot E)=(J_{\rm e} \cdot E)_{\rm Non ohmic} + (J_{\rm e} \cdot E)_{\rm ohmic}
\end{align}
where
\begin{align}\label{split2}
(J_{\rm e} \cdot E)_{\rm Non ohmic} &= \text{Term I} + \text{Term II} + \text{Term III} \nonumber \\ &+ \text{Term IV} + \text{Term V} \\ \label{split3} (J_{\rm e} \cdot E)_{\rm ohmic} &= \text{Term VI}
\end{align}
In turn the Non ohmic contribution can be split up as follows \cite{lafleur14:035010}
\begin{align}\label{split4}
(J_{\rm e} \cdot E)_{\rm Non ohmic} = (J_{\rm e} \cdot E)_{\rm Inertia} + (J_{\rm e} \cdot E)_{\rm Pressure}
\end{align}
where
\begin{align}\label{split5}
(J_{\rm e} \cdot E)_{\rm Inertia} &= \text{Term I} + \text{Term II} + \text{Term III} \\ \label{split6}
(J_{\rm e} \cdot E)_{\rm Pressure} &= \text{Term IV} + \text{Term V}
\end{align}
This will be useful later when we identify the different contributions to the electron power absorption. We underline that the same split applied to the electron power absorption can be applied to the electric field formula shown in Eq.~\eqref{equation4}.
\section{Results and Discussion}\label{RaD}
The sheath location is determined by assuming that the density of negatively charged species has fallen to half the density of the positive charged species. In more detail the sheath edge position, which is defined as $x_{\rm sh}(t)$, is taken to be the position $x$, where the condition $n_{\rm e}(x,t)/n_{\rm i} = 1/2$ is satisfied.  This determines the location of the grey shadowed rectangles hiding the sheath regions in the various plots shown in this section. We hide the sheath regions directly adjacent to the electrodes as the electron density is very low. 
\begin{figure}[!ht]
\begin{center}
\includegraphics[width=8 cm]{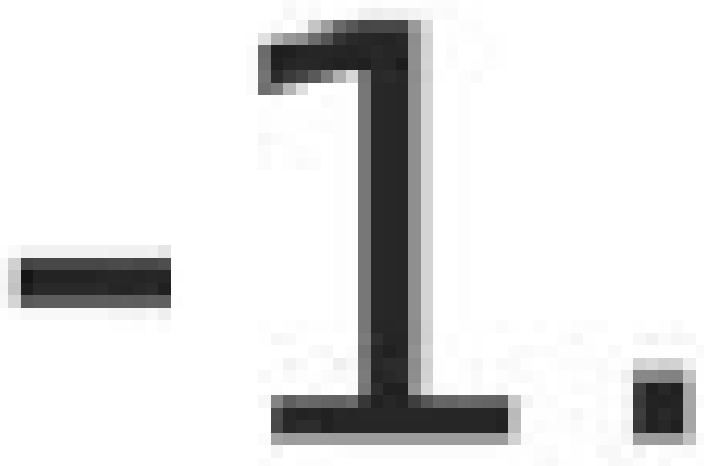}   
\includegraphics[width=8 cm]{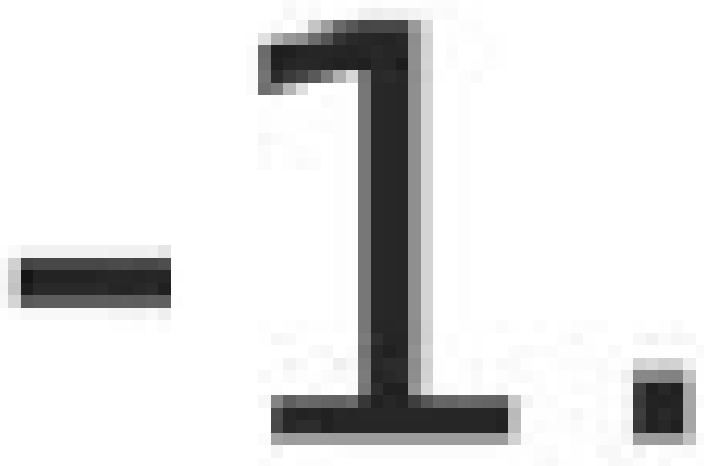}   
\includegraphics[width=8 cm]{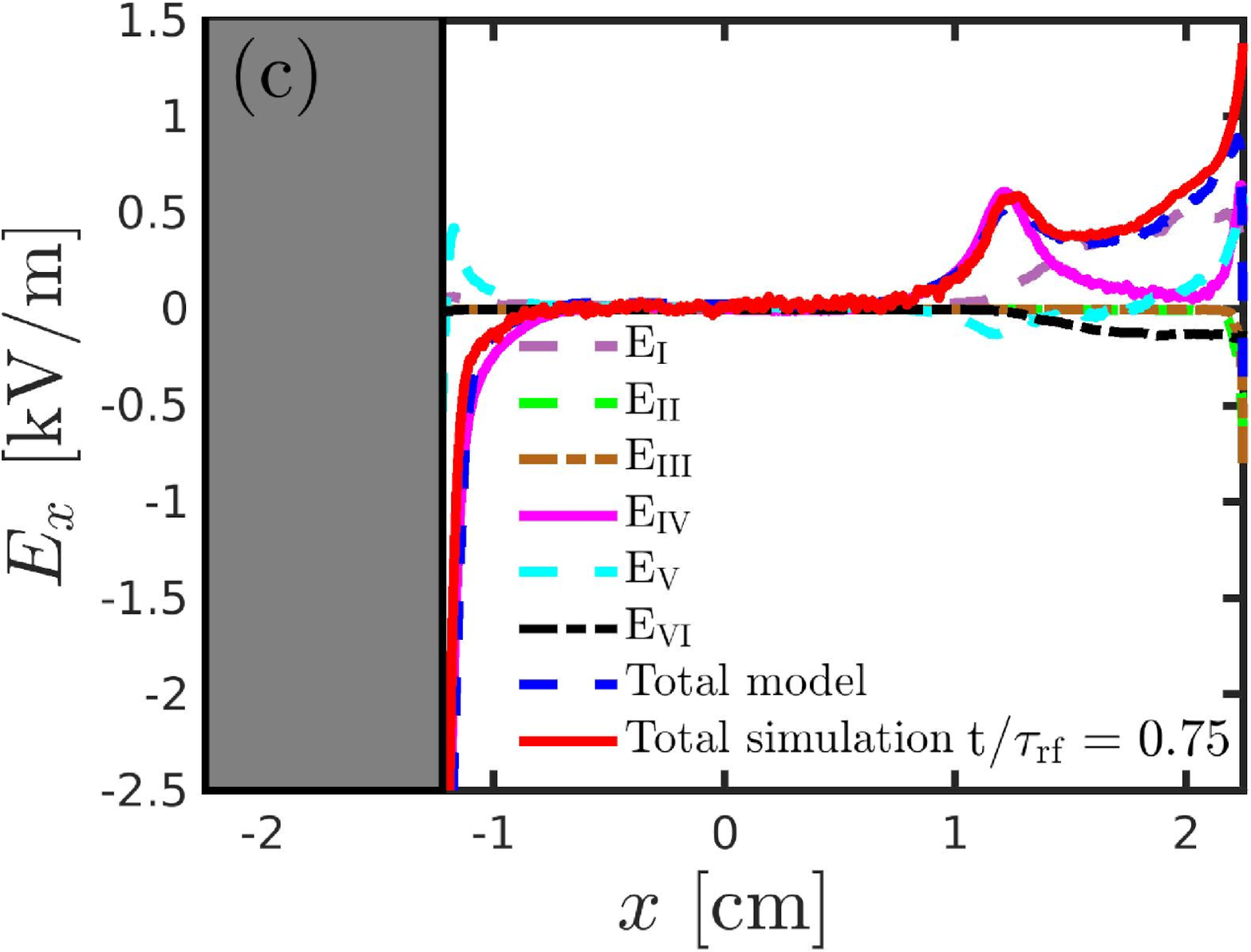}   
\end{center}
\caption{\label{fig2}The electric field profile of the terms that constitute Eq.~\eqref{equation4} and their summation compared with the result of the simulations at (a) $t / \tau_{\rm rf}=0.25$ from the left electrode to the right sheath edge, at (b) $t / \tau_{\rm rf}=0.50$ from the left to the right sheath edge, at (c) $t / \tau_{\rm rf}=0.75$ from the left sheath edge to the right electrode, for a parallel plate capacitively coupled oxygen discharge at 100 mTorr for 45 mm of gap separation driven by a 400 V voltage source at driving frequency of 13.56 MHz.}
\end{figure}
Figure \ref{fig2} shows the electric field profile of the terms that constitute Eq.~\eqref{equation4} and their summation compared with the results of simulations at $t / \tau_{\rm rf}=0.25$ from the left electrode to the right sheath edge, at $t / \tau_{\rm rf}=0.50$ from the left to the right sheath edge, at $t / \tau_{\rm rf}=0.75$ from the left sheath edge to the right electrode. For all the three time slices an almost perfect match between the overall terms summation and the result from the simulations is observed. 

At $t / \tau_{\rm rf}=0.25$ (Figure \ref{fig2} (a)) we see that the contribution from Terms II and III is negligible. On the other hand, the main contribution comes from Terms I, IV, V and VI. Term I is almost zero within the bulk region up to the right sheath edge while it decreases to negative values as it approaches the left sheath edge up to the left electrode. The electric field inertia term due to temporally varying electron velocity becomes negative approaching the left electrode, indicating that the electron velocity gradient is positive near the left electrode (Eq.~\eqref{equation4}). Term IV is flat and zero within the bulk region and sharply increases (sharply decreases) while approaching the right (left) sheath edge. An absolute minimum in the Term IV profile is observed on the sheath side of the left sheath edge. Moreover, once passed this minimum, the profile of Term IV is approximatively constant over the left sheath region and it decreases while approaching the left electrode. Term V is flat and zero over the full bulk gap length. A small local minimum (maximum) in the Term V profile is observed on the sheath side of the right (left) sheath edge. Then it steeply increases (sharply decreases) while approaching the sheath side of the right sheath edge (the left electrode). Term VI is flat and zero within the bulk region up to the right sheath edge while it slightly increases as it approaches the left sheath keeping an almost constant value while approaching the left electrode. Finally, Figure \ref{fig2} (a) shows that the only important contributions to the electric field at $t / \tau_{\rm rf}=0.25$ comes from the inertia term related to the temporal gradient of the electron velocity (Term I), from the pressure gradient related terms (Term IV and V) and from the ohmic heating term (Term VI). At $t / \tau_{\rm rf}=0.50$ (Figure \ref{fig2} (b)) we see that the contribution from Terms I and III are negligible and Term II is small except near the sheath edges. Term IV is flat and zero within the bulk region and it sharply increases (steeply decreases) as it approaches the bulk side of the right (left) sheath edge. Term V is flat and zero within the bulk region and slightly decreases (sharply increases) as it approaches the sheath side of the right (left) sheath edge. Term VI is flat and zero within the bulk region up to both the sheath edges and it sharply decreases as it approaches the bulk side of both the sheath edges. Figure \ref{fig2} (b) shows that the only important contributions to the electric field at $t / \tau_{\rm rf}=0.50$ comes from the pressure gradient related terms (Terms IV and V) and from the ohmic contribution (Term VI). At $t / \tau_{\rm rf}=0.75$ (Figure \ref{fig2} (c)) we see almost a mirror image of Figure \ref{fig2} (a). The contributions from Terms II and III are negligible and Term VI is small. The  main contribution comes from Terms I, IV, and V. The only significant contributions to the electric field at $t / \tau_{\rm rf}=0.75$ comes from the inertia term related to the temporal gradient of the electron velocity and from the pressure gradient related terms. 
Therefore, the most significant contribution to the electric field within the bulk plasma at 100 mTorr is due to the pressure gradient terms. An almost perfect match between the calculated electric field profile and the result from simulation can be observed for all the time steps considered as shown in Figures \ref{fig2} (a), (b) and (c).
\begin{figure}[!ht]
\begin{center}
\includegraphics[width=8 cm]{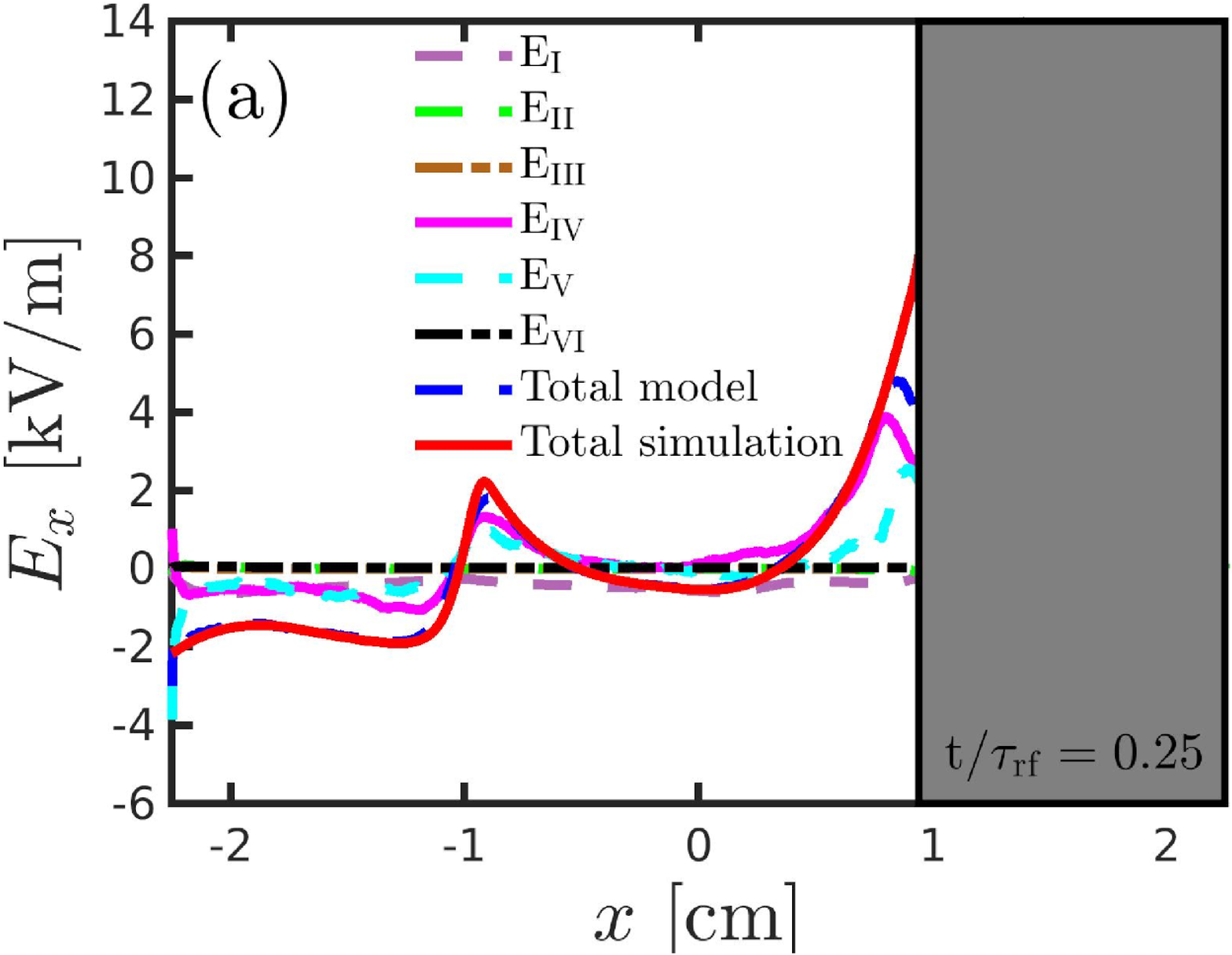}   
\includegraphics[width=8 cm]{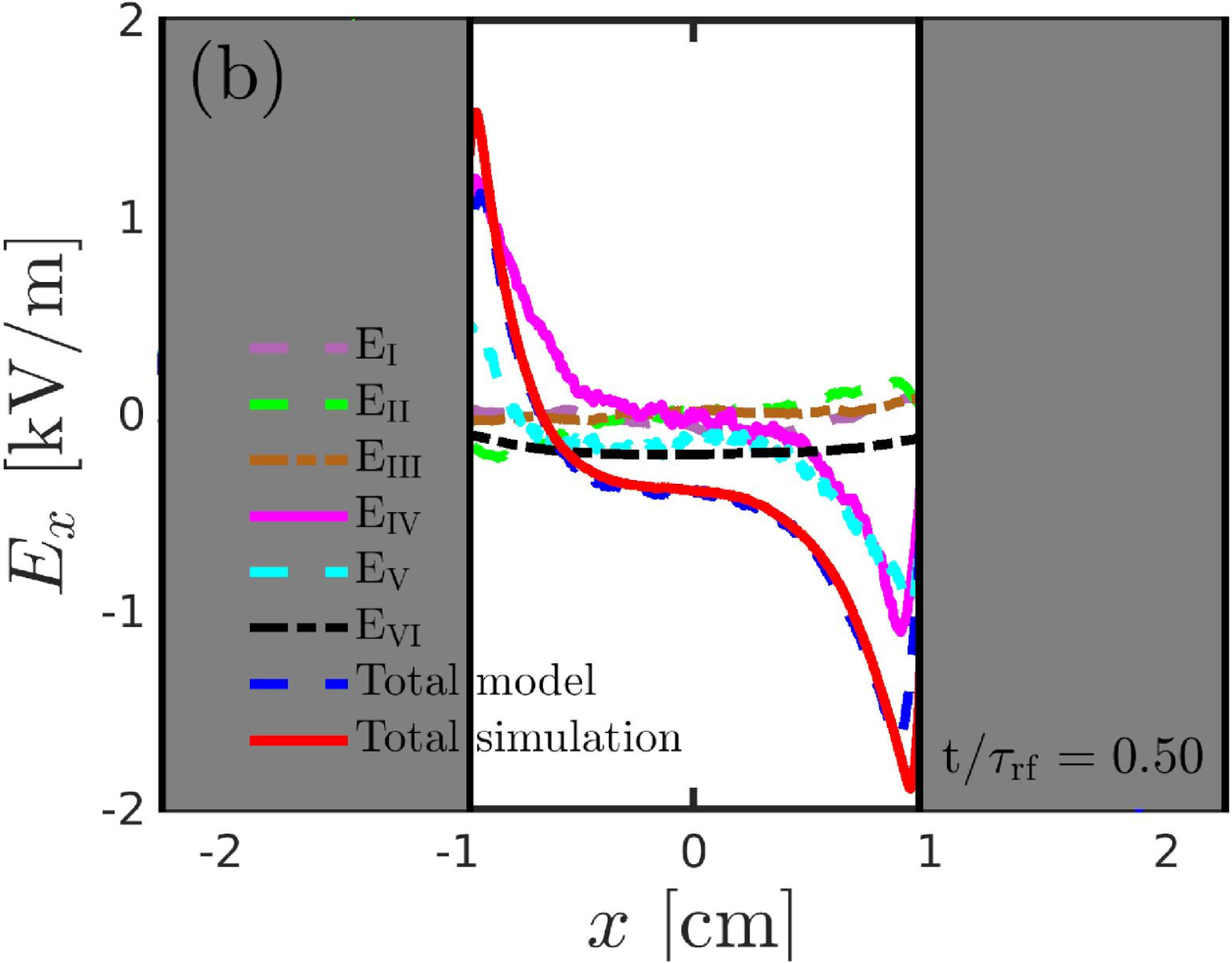}   
\includegraphics[width=8 cm]{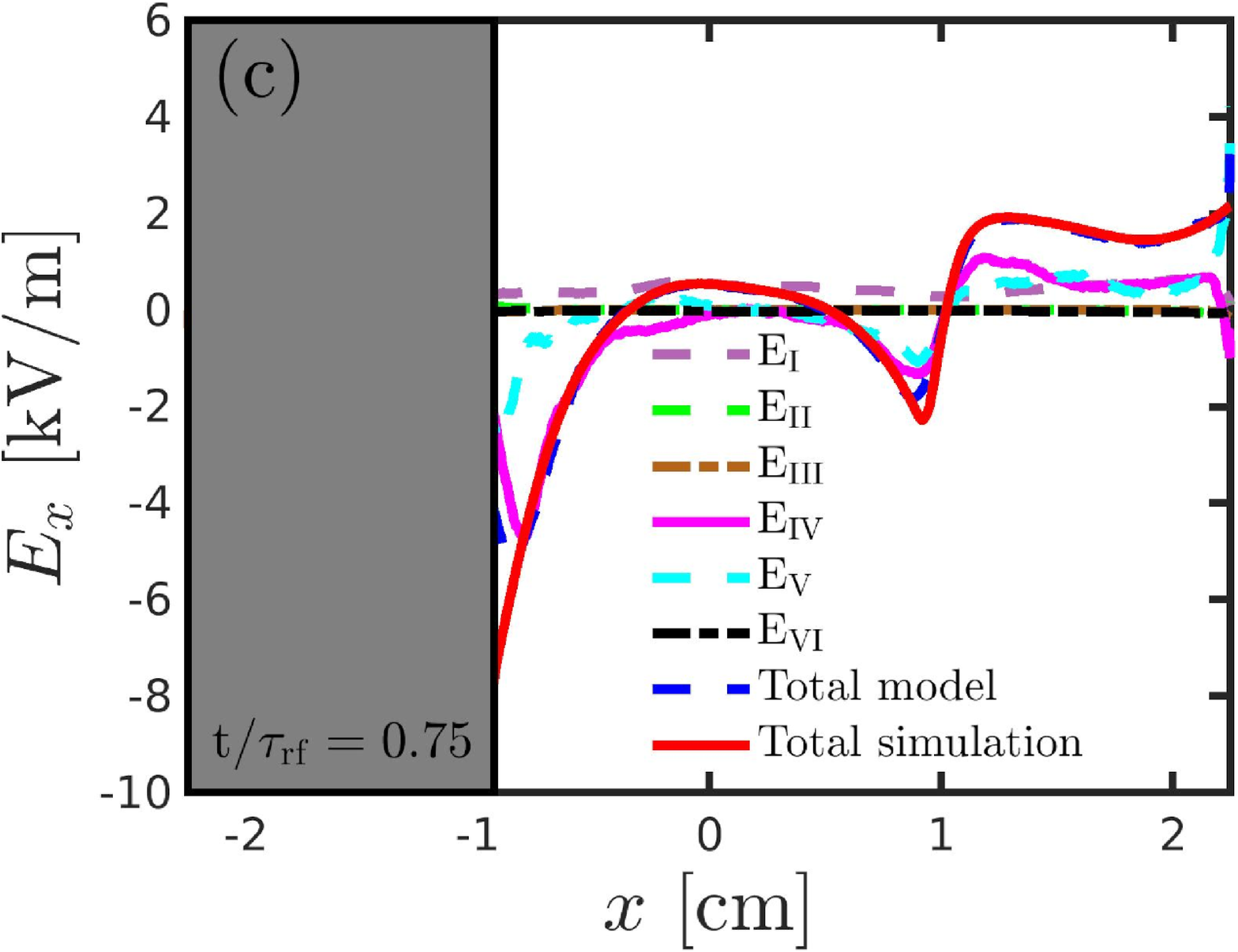}   
\end{center}
\caption{\label{fig2b}The electric field profile of the terms that constitute Eq.~\eqref{equation4} and their summation compared with the result of the simulations at (a) $t / \tau_{\rm rf}=0.25$ from the left electrode to the right sheath edge, at (b) $t / \tau_{\rm rf}=0.50$ from the left to the right sheath edge, at (c) $t / \tau_{\rm rf}=0.75$ from the left sheath edge to the right electrode (c), for a parallel plate capacitively coupled oxygen discharge at 10 mTorr for 45 mm of gap separation driven by a 400 V voltage source at driving frequency of 13.56 MHz.}
\end{figure}

Figure \ref{fig2b} shows the electric field profile at 10 mTorr of the terms that constitute Eq.~\eqref{equation4} and their summation compared with the result of the simulations at (a) $t / \tau_{\rm rf}=0.25$ from the left electrode to the right sheath edge, at (b) $t / \tau_{\rm rf}=0.50$ from the left to the right sheath edge, at (c) $t / \tau_{\rm rf}=0.75$ from the left sheath edge to the right electrode. At $t / \tau_{\rm rf}=0.25$ (Fig.~\ref{fig2b} (a)) we see that the contribution from terms II and VI is negligible while Term III is small. We observe that the main contribution to the electric field comes from terms IV and V. Term IV is zero within the discharge center and approximatively flat and zero within the inner bulk region and sharply increases while approaching the bulk side of both the sheath edges. A lower (higher) maximum on the bulk side of the left (right) sheath edge is observed. On the other hand, Term V has a similar behaviour except that it increases less steeply while approaching the bulk side of the right sheath edge.  The local maximum in Term V on the left sheath edge overlaps almost perfectly with the local maximum in Term IV,  total terms summation and the result from the simulations placed in the same location as well as the respective profiles within the inner bulk region. Finally, Figure \ref{fig2b} (a) shows that the only important contributions to the electric field at $t / \tau_{\rm rf}=0.25$ comes from the pressure gradient related terms (Terms IV and V). At $t / \tau_{\rm rf}=0.50$ (Fig.~\ref{fig2b} (b)) we see that the contribution from Term I is negligible and Terms II,  III and VI are small. We also observe that the main contribution to the electric field comes from terms IV and V. Term IV is flat and zero within the inner core of the plasma bulk and decreases (increases) while approaching the bulk side of the right (left) sheath edge. Then it increases again once passed the right sheath edge building an absolute minimum. Finally, Figure \ref{fig2b} (b) shows that the only important contributions to the electric field at
$t / \tau_{\rm rf}=0.50$ come from the pressure gradient related terms (Terms IV and V). Moreover, we observe that in the $t / \tau_{\rm rf}=0.50$ case Terms II, IV and V share the same importance in contribution to the electric field at both 10 and 100 mTorr while the contribution from Term III (Term VI) is lacking at 100 mTorr (10 mTorr). At $t / \tau_{\rm rf}=0.75$ (Fig.~\ref{fig2b} (c)) we see  a mirror image of the  $t / \tau_{\rm rf}=0.25$ case. Terms II,  III and VI are negligible and Term I is small. The main contribution to the electric field comes from terms IV and V. The local minimum in Term V on the right sheath edge overlaps almost perfectly with the local minimum in Term IV, total terms summation and the result from the simulations placed in the same location as well as the respective profiles within the inner bulk region. To summarize the only important contributions to the electric field at $t / \tau_{\rm rf}=0.75$ comes from the pressure gradient related terms (Terms IV and V).

An almost perfect match between the calculated electric field profile and the result from simulation can be observed over the full gap length up to the bulk side of the expanding sheath edge for all the three time slices considered. Moreover, at $t / \tau_{\rm rf}=0.25$ and $t / \tau_{\rm rf}=0.75$, the calculated electric field sharply understimates the electric field coming from the simulations while approaching the bulk side of the fully collapsed sheath edge, while at $t / \tau_{\rm rf}=0.50$ the difference is very small.
 We observe that the inertia term (Term I), for all the three time slices considered, is negligible compared to the 100 mTorr case (Figure \ref{fig2}). Such a term is absent due to the presence of the negative charged stripes placed on bulk side of the collapsing sheath edge, which prevents the electrons from crossing the collapsing sheath edge (see discussion in Section III). Moreover, due to the presence of the positive charged stripes placed on the sheath side over the full rf period, the electrons are prevented from increasing their own velocity. Since at 10 mTorr there is higher number of electrons within the collapsing sheath region than at 100 mTorr (Figure \ref{figA} (a) and Figure \ref{figA2} (a) respectively), the displacement current is lower and the temporal change in the electron velocity due to the time varying electric field is also lower.
\begin{figure}[!ht]
\begin{center}
\includegraphics[width=8 cm]{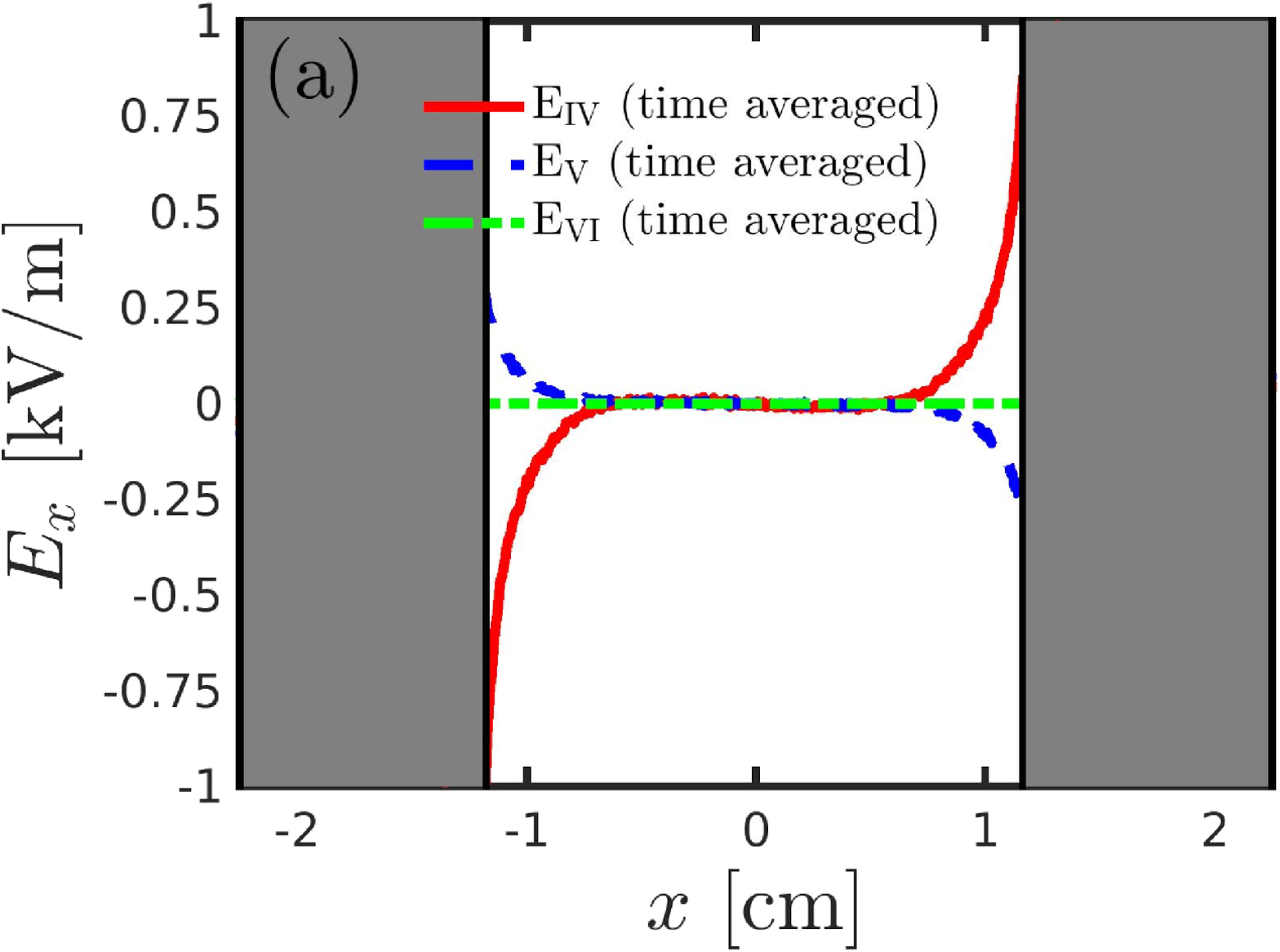}   
\includegraphics[width=8 cm]{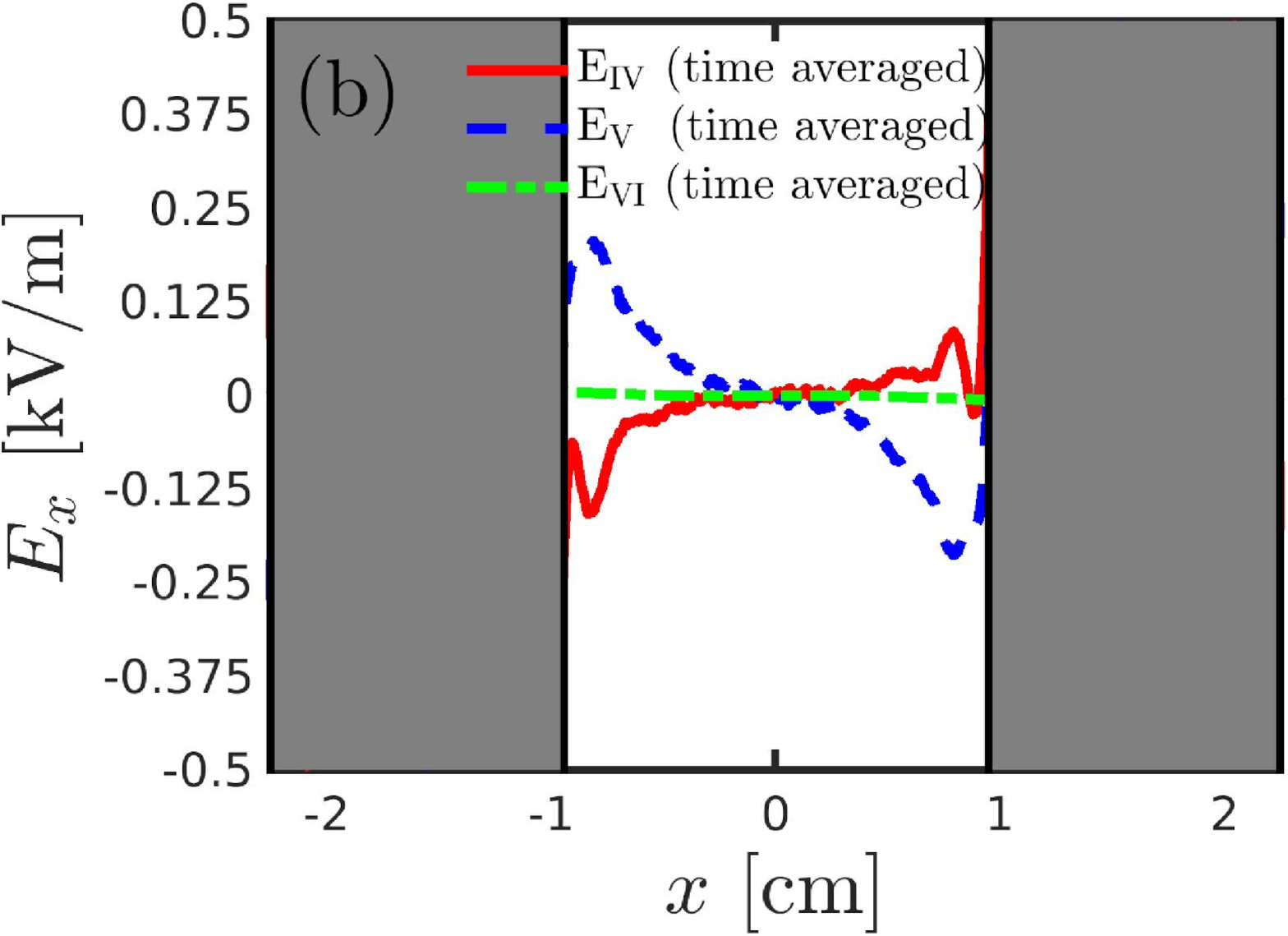}   
\end{center}
\caption{\label{fig3}The time averaged electric field profile at (a) 100 mTorr and at (b) 10 mTorr of Term IV (red line), Term V (blue dashed line), Term VI (green dotted dashed line) from Eq.~\eqref{equation4} from the left to the right sheath edge for a parallel plate capacitively coupled oxygen discharge for 45 mm of gap separation driven by a 400 V voltage source at driving frequency of 13.56 MHz.}
\end{figure}

Figure \ref{fig3} (a) shows the time averaged electric field profile at 100 mTorr for the three main contributing terms to the calculated electric field using Eq.~\eqref{equation4} from the left to the right (time averaged) sheath edge: Term IV (red line), V (blue dashed line) and VI (green dotted dashed line). We see that all the three terms considered are flat and zero within the bulk region. Term IV steeply increases (decreases) while approaching the right (left) sheath edge. On the other hand, Term V sharply increases (decreases) while approaching the sheath side of the left (right) edge. Finally, Figure \ref{fig3} (a) shows that the only important contribution to the time averaged electric field comes from the pressure terms (Terms IV and V).

Figure \ref{fig3} (b) shows the time averaged electric field profile in a discharge operated at 10 mTorr for the three main contributing terms to the calculated electric field using Eq.~\eqref{equation4} from the left to the right (time averaged) sheath edge: Term IV (red line), V (blue dashed line) and VI (green dotted dashed line). We see that all the three terms considered are zero within the discharge center. Term IV sharply increases (decreases) while approaching the bulk side of the right (left) sheath edge. Term V  steeply decreases (increases) while approaching the bulk side of the right (left) sheath edge and it steeply increases (decreases) while crossing the right (left) sheath edge. Finally, Figure \ref{fig3} (b) shows that the only important contribution to the time averaged electric field comes from the pressure terms (Terms IV and V), just like at 100 mTorr.
\begin{figure}[!ht]
\begin{center}
\includegraphics[width=8 cm]{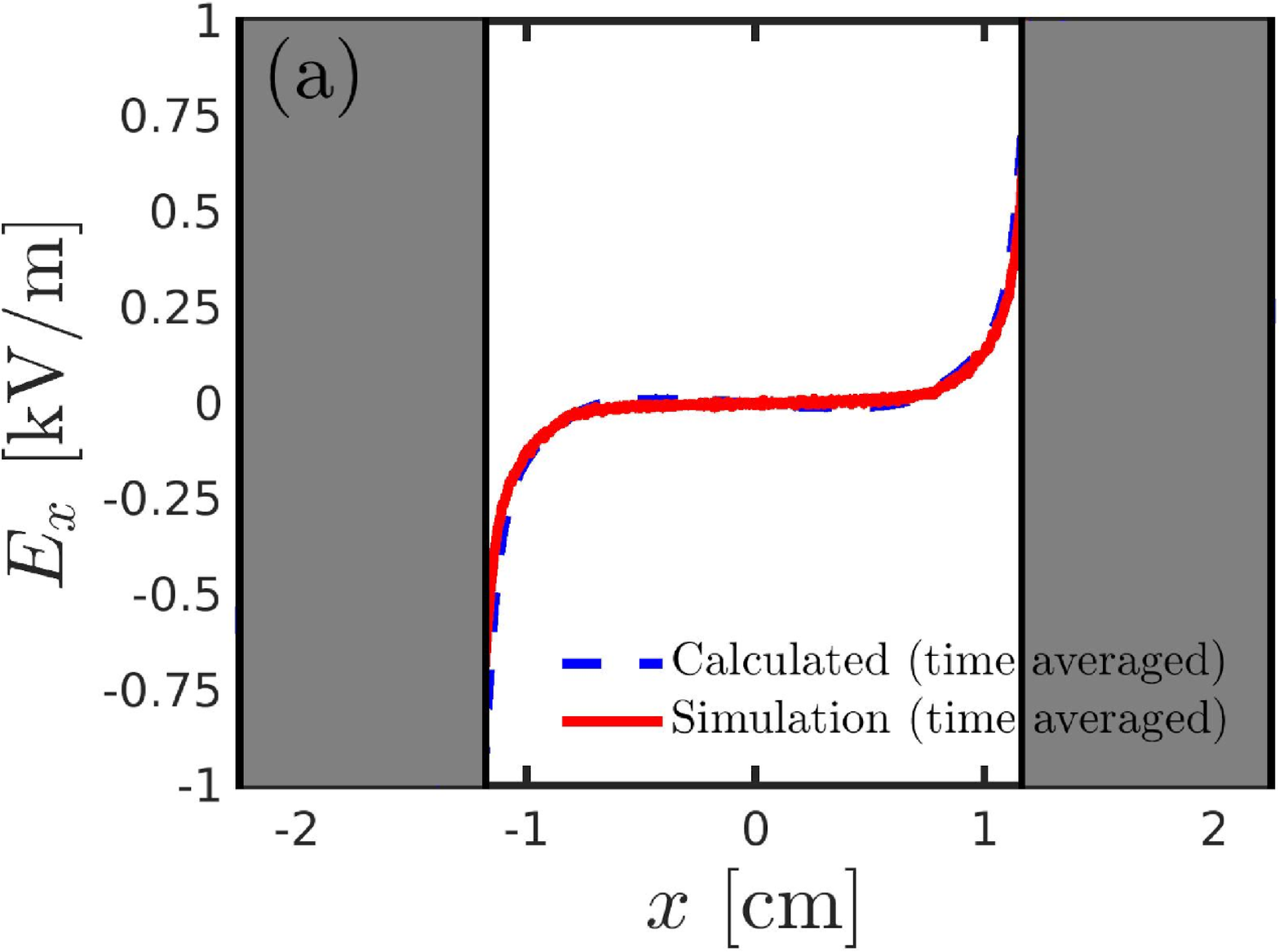}
\includegraphics[width=8 cm]{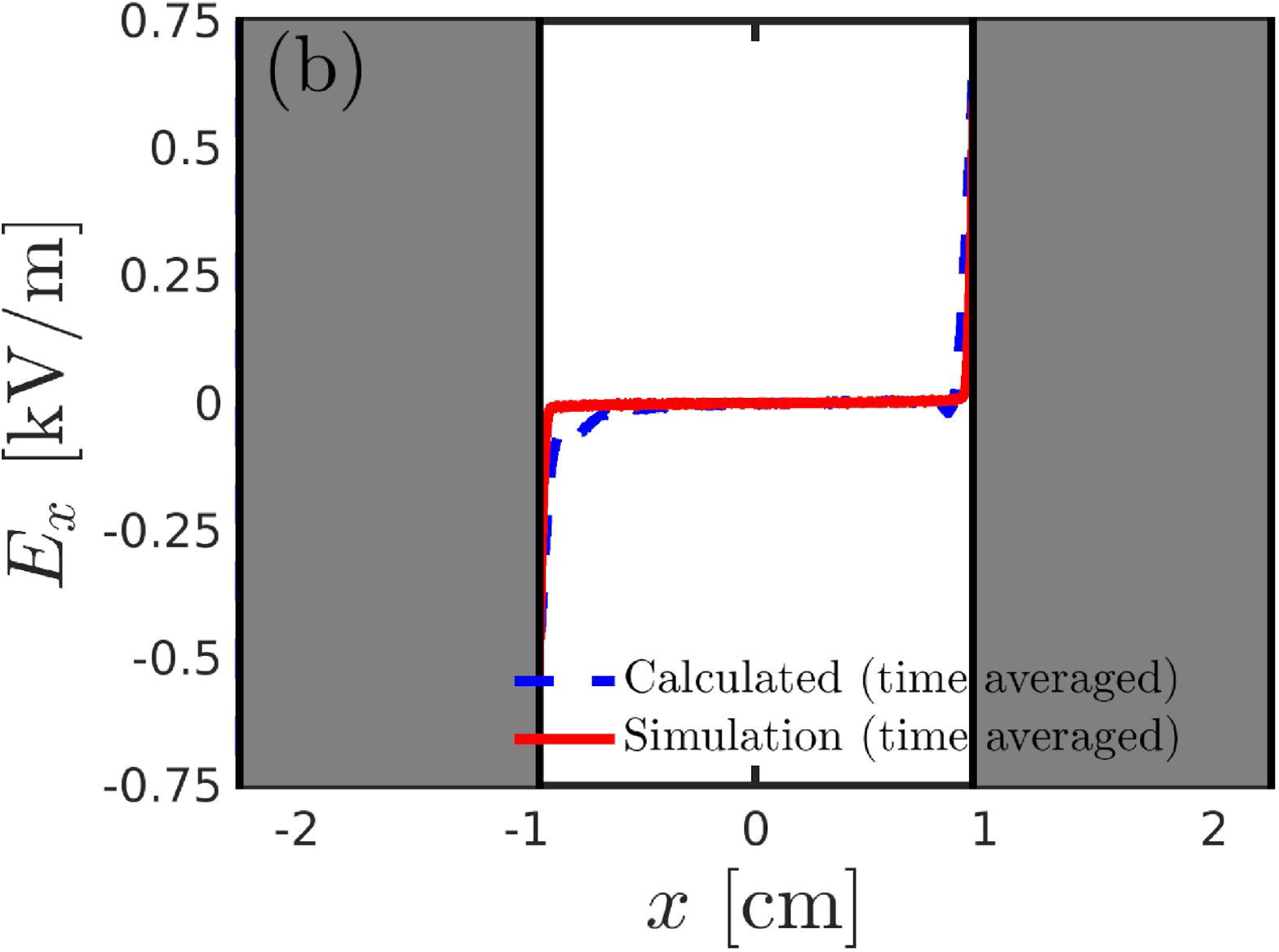}   
\end{center}
\caption{\label{fig5}The time averaged electric field profile calculated using Eq.~\eqref{equation4} (blue dashed line) and the result from simulations (red line) from the left to the right sheath edge at (a) 100 mTorr and at (b) 10 mTorr for a parallel plate capacitively coupled oxygen discharge for 45 mm of gap separation driven by a 400 V voltage source at driving frequency of 13.56 MHz.}
\end{figure}

Figure \ref{fig5} (a) shows the time averaged electric field at 100 mTorr calculated using Eq.~\eqref{equation4} (blue dashed line) and the result from simulations (red line) from the left to the right sheath edge. An almost perfect match is observed over the gap length considered. Figure \ref{fig5} (b) shows the time averaged electric field at 10 mTorr calculated using Eq.~\eqref{equation4} (blue dashed line) and the result from simulations (red line) from the left to the right sheath edge. We observe an almost perfect match within the inner bulk gap length. However, the calculated electric field overstimates (understimates) the electric field from the simulation on the bulk side of both the sheath edges,  resembling the observed slight difference between the calculated and simulated electric field at different time steps on the bulk side of the expanding sheath edge (Figure \ref{fig2b}).
\begin{figure}[!ht]
\begin{center}
\includegraphics[width=8 cm]{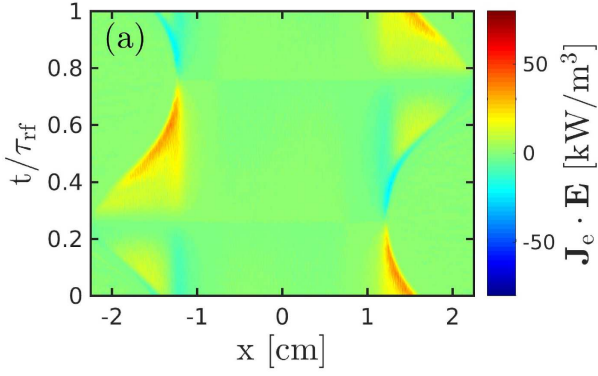}   
\includegraphics[width=8 cm]{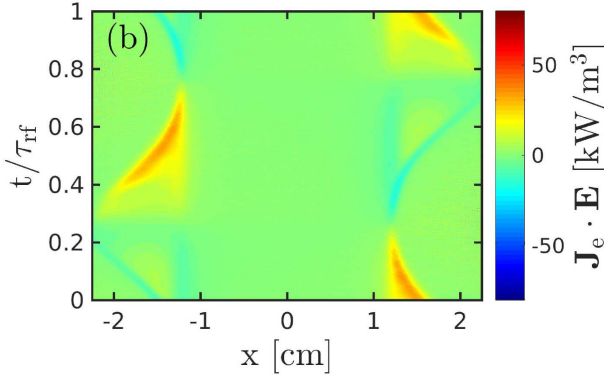}   
\end{center}
\caption{\label{fig6}The spatio-temporal behaviour of the electron power absorption over the full gap length (a) calculated using Eq.~\eqref{equation5} and (b) the result from simulation for a parallel plate capacitively coupled oxygen discharge at 100 mTorr for 45 mm of gap separation driven by a 400 V voltage source at driving frequency of 13.56 MHz.}
\end{figure}

Figure \ref{fig6} shows the  spatio-temporal behavior of the electron power absorption
${\bf J}_{\rm e} \cdot {\bf E}$ at 100 mTorr. The figures show the electron power absorption calculated using Eq.~\eqref{equation5} (Figure \ref{fig6} (a)) and from the simulation over the full gap length (Figure \ref{fig6} (b)). The ordinate covers the full rf cycle. We see that almost all the electron power absorption occurs during the sheath expansion and that the electron power absorption mode is a pure $\alpha$-mode. The calculated electron power absorption almost perfectly matches the result from the simulation on both the sheath edges. We recall that the theoretical model doesn't take the sheath dynamics into account, since it has been built for the bulk and collapsed sheath region only. Within the gap length we see an almost perfect match between the electron power absorption coming from the theoretical model and the simulation.
\begin{figure}[!ht]
\begin{center}
\includegraphics[width=8 cm]{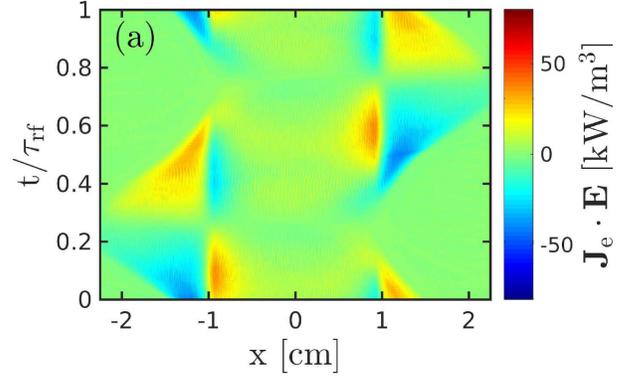}   
\includegraphics[width=8 cm]{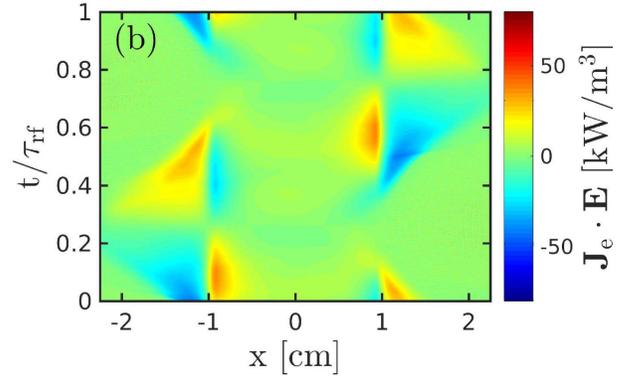}   
\end{center}
\caption{\label{fig6b}The spatio-temporal behaviour of the electron power absorption over the full gap length calculated (a) using Eq.~\eqref{equation5} and (b) the result from simulation for a parallel plate capacitively coupled oxygen discharge at 10 mTorr for 45 mm of gap separation driven by a 400 V voltage source at driving frequency of 13.56 MHz.}
\end{figure}

Figure \ref{fig6b} shows the  spatio-temporal behavior of the electron power absorption
${\bf J}_{\rm e} \cdot {\bf E}$ at 10 mTorr, where ${\bf J}_{\rm e}$ and ${\bf E}$ are the spatially and temporally varying electron current
density and electric field, respectively. The figures show the electron power absorption for the theoretical model (Eq.~\eqref{equation5}) (Figure \ref{fig6b} (a)) and from the simulation results over the full gap length (Figure \ref{fig6b} (b)). The ordinate covers the full rf cycle. Firstly, we observe that at 10 mTorr a significant power absorption (red and yellow areas) and some electron cooling (dark blue areas) are evident within the plasma bulk region. The electron power absorption appears during the sheath expansion (collapse) on the sheath side (on the bulk side) of the sheath edge, while there is electron cooling during the sheath expansion (collapse) on the bulk side of the sheath edge (on the electrode side).
Indeed, at 10 mTorr the electron heating mechanism is a combination of a drift ambipolar (DA) heating in the bulk plasma and stochastic heating due to the sheath oscillation ($\alpha$-mode), as it has been shown in our previous works \citep{proto18:074002, proto18:65, gudmundsson19:045012}. Secondly, we observe that the calculated electron power absorption resembles well the result from the simulation which is slightly overstimated within the sheath region only.
\begin{figure}[!ht]
\begin{center}
\includegraphics[width=8 cm]{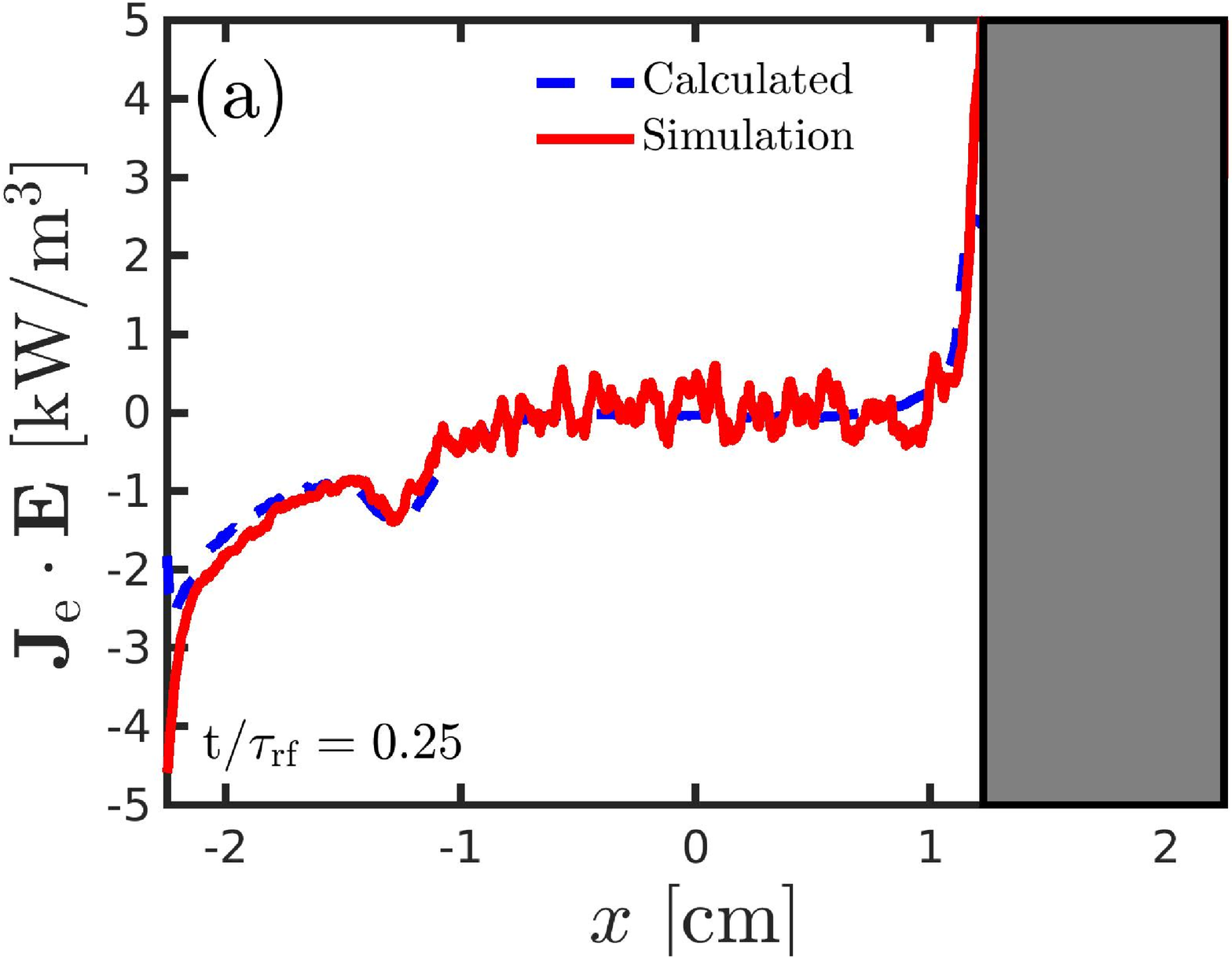}   
\includegraphics[width=8 cm]{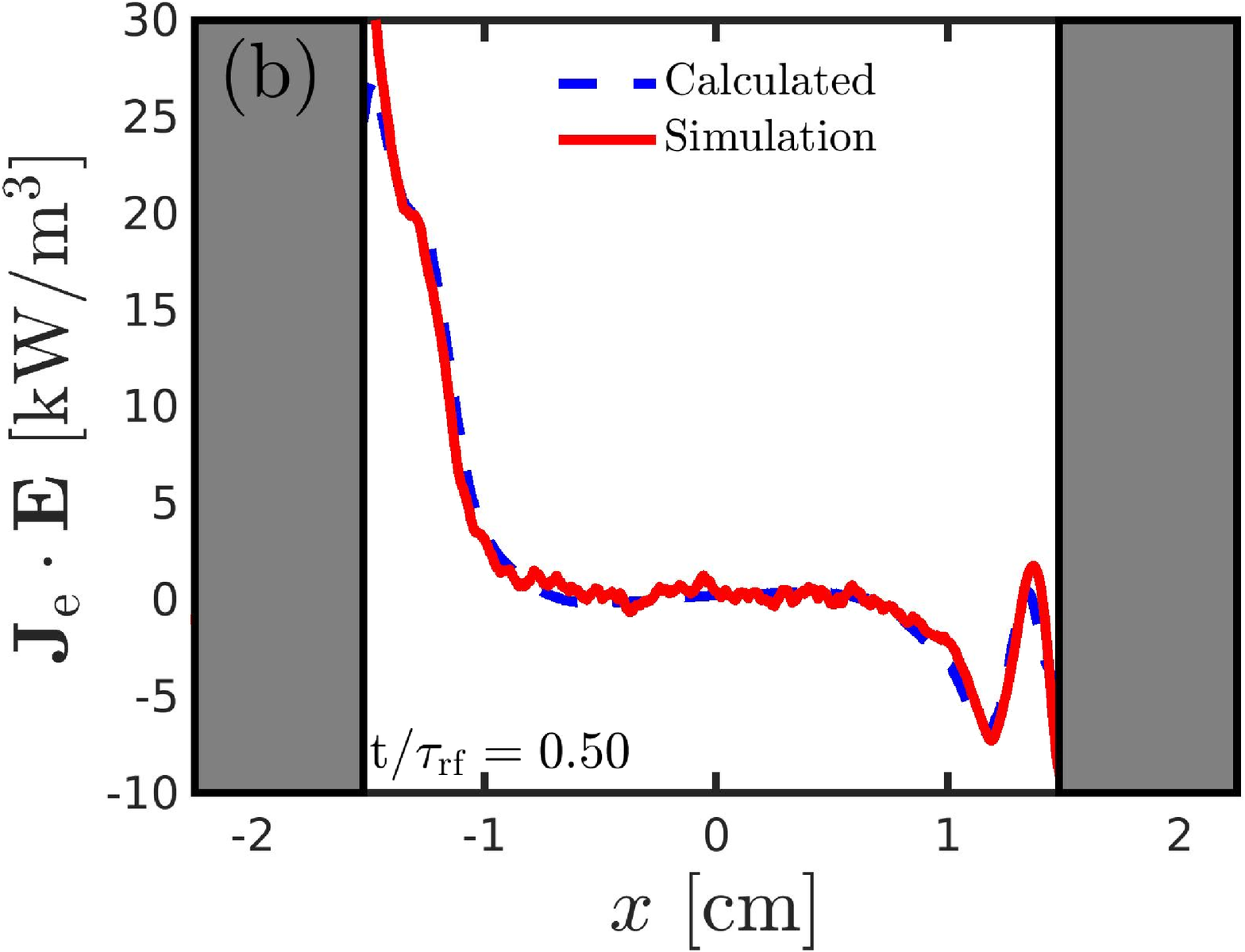}   
\includegraphics[width=8 cm]{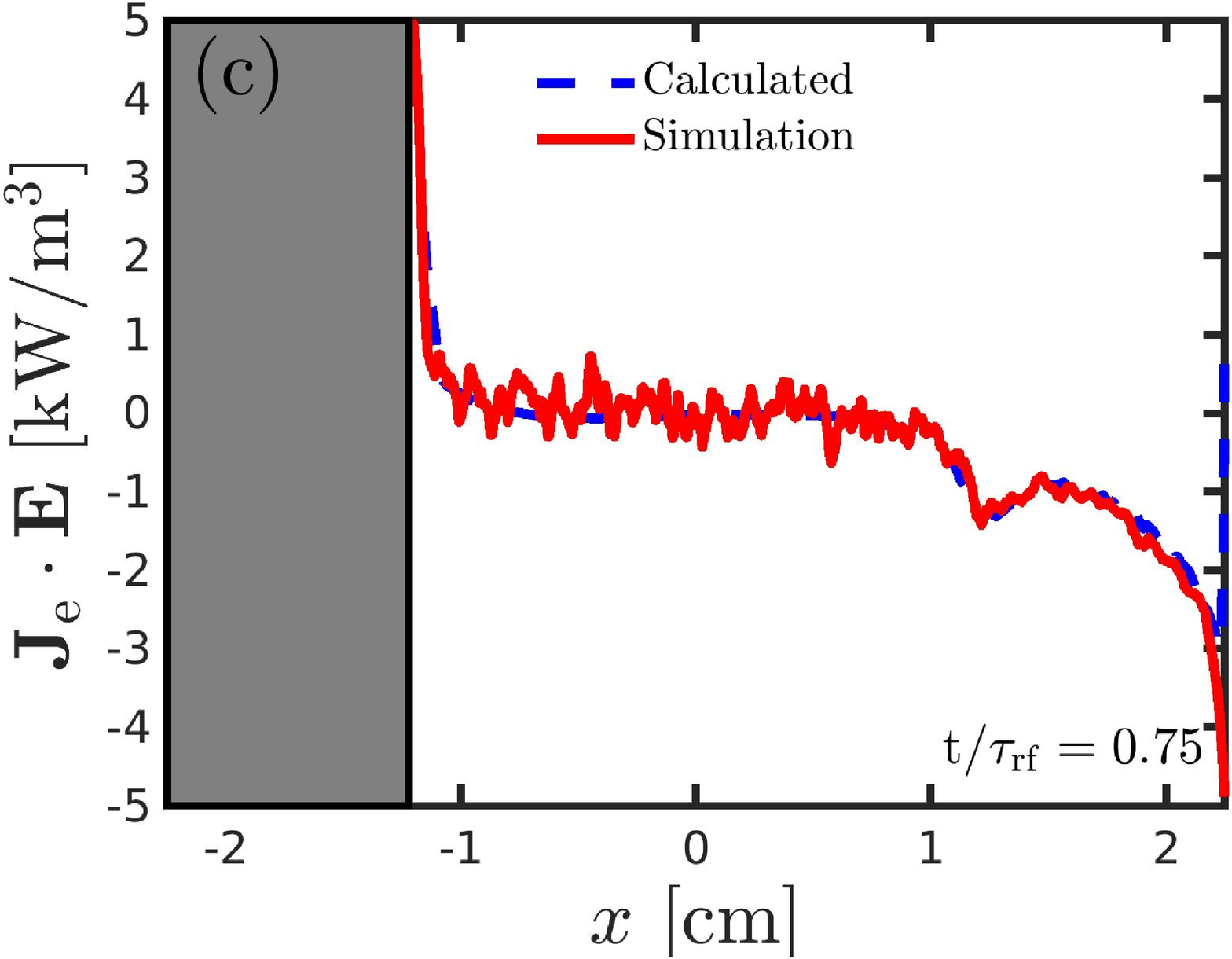}   
\end{center}
\caption{\label{fig8}The electron power absorption profile from Eq.~\eqref{equation5} (blue dashed line) at (a) $t / \tau_{\rm rf}=0.25$ and the result from simulations (red line) from the left electrode to the right sheath edge, at (b) $t / \tau_{\rm rf}=0.50$ from the left to the right sheath edge, at (c) $t / \tau_{\rm rf}=0.75$ from the left sheath edge to the right electrode, for a parallel plate capacitively coupled oxygen discharge at 100 mTorr for 45 mm of gap separation driven by a 400 V voltage source at driving frequency of 13.56 MHz.}
\end{figure}

Figure \ref{fig8} shows the electron power absorption profile at 100 mTorr calculated using Eq.~\eqref{equation5} (blue dashed line) and the result from simulations (red line) at (a) $t / \tau_{\rm rf}=0.25$ from the left electrode to the right sheath edge, at (b) $t / \tau_{\rm rf}=0.50$ from the left to the right sheath edge, at (c) $t / \tau_{\rm rf}=0.75$ from the left sheath edge to the right electrode. An almost perfect match is observed for all the time steps considered as shown in Figures \ref{fig8} (a), (b) and (c).
\begin{figure}[!ht]
\begin{center}
\includegraphics[width=8 cm]{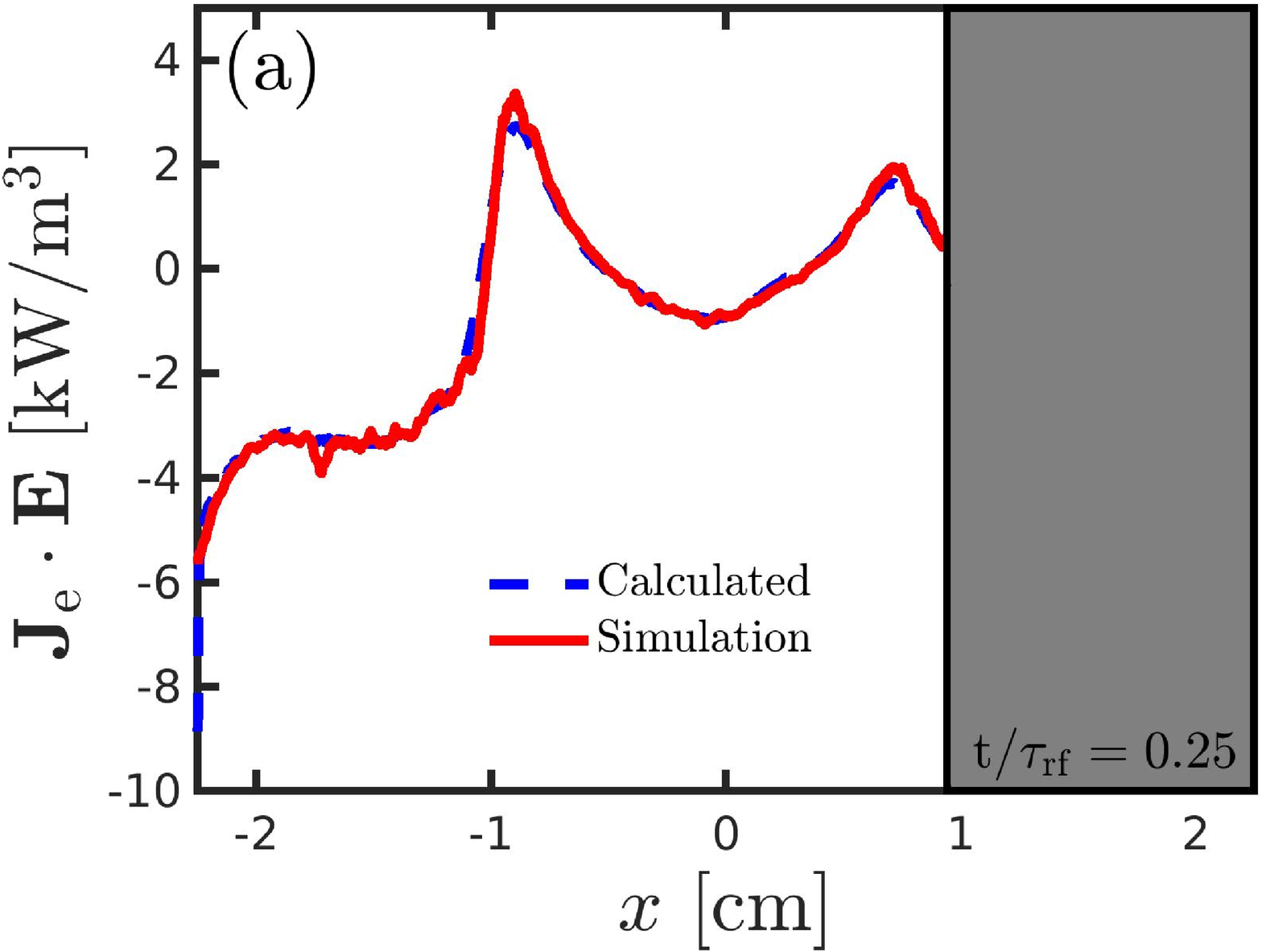}   
\includegraphics[width=8 cm]{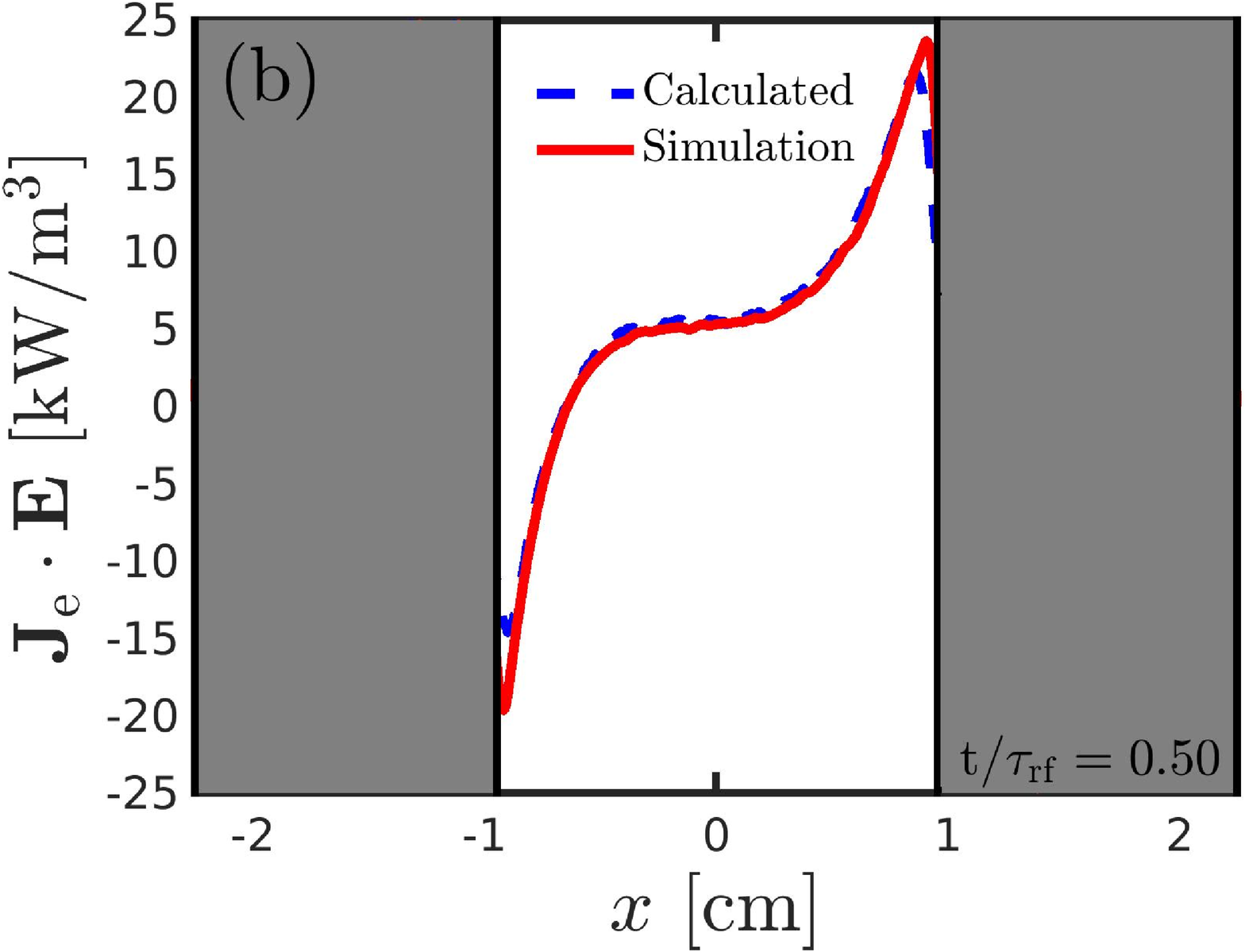}   
\includegraphics[width=8 cm]{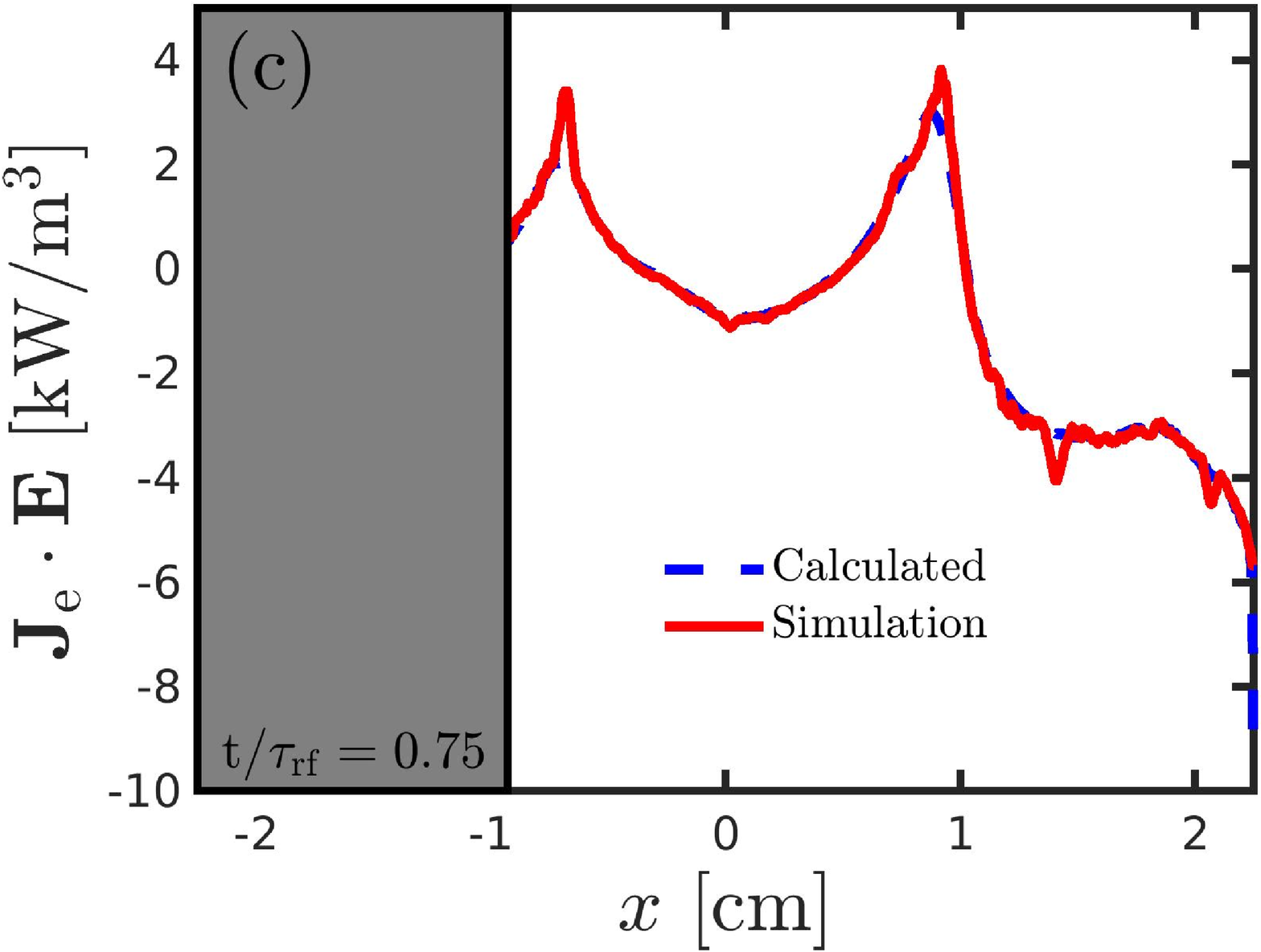}   
\end{center}
\caption{\label{fig8b}The electron power absorption profile from Eq.~\eqref{equation5} (blue dashed line) at (a) $t / \tau_{\rm rf}=0.25$ and the result from simulations (red line) from the left electrode to the right sheath edge, at (b) $t / \tau_{\rm rf}=0.50$ from the left to the right sheath edge, at (c) $t / \tau_{\rm rf}=0.75$ from the left sheath edge to the right electrode, for a parallel plate capacitively coupled oxygen discharge at 10 mTorr for 45 mm of gap separation driven by a 400 V voltage source at driving frequency of 13.56 MHz.}
\end{figure}

Figure \ref{fig8b} shows the electron power absorption profile at 10 mTorr calculated using Eq.~\eqref{equation5} (blue dashed line) and the result from simulations (red line) at $t / \tau_{\rm rf}=0.25$ from the left electrode to the right sheath edge (a), at $t / \tau_{\rm rf}=0.50$ from the left to the right sheath edge (b), at $t / \tau_{\rm rf}=0.75$ from the left sheath edge to the right electrode (c). An almost perfect match is observed for all the time steps considered as shown in Figures \ref{fig8b} (a), (b) and (c).
\begin{figure}[!ht]
\begin{center}
\includegraphics[width=8 cm]{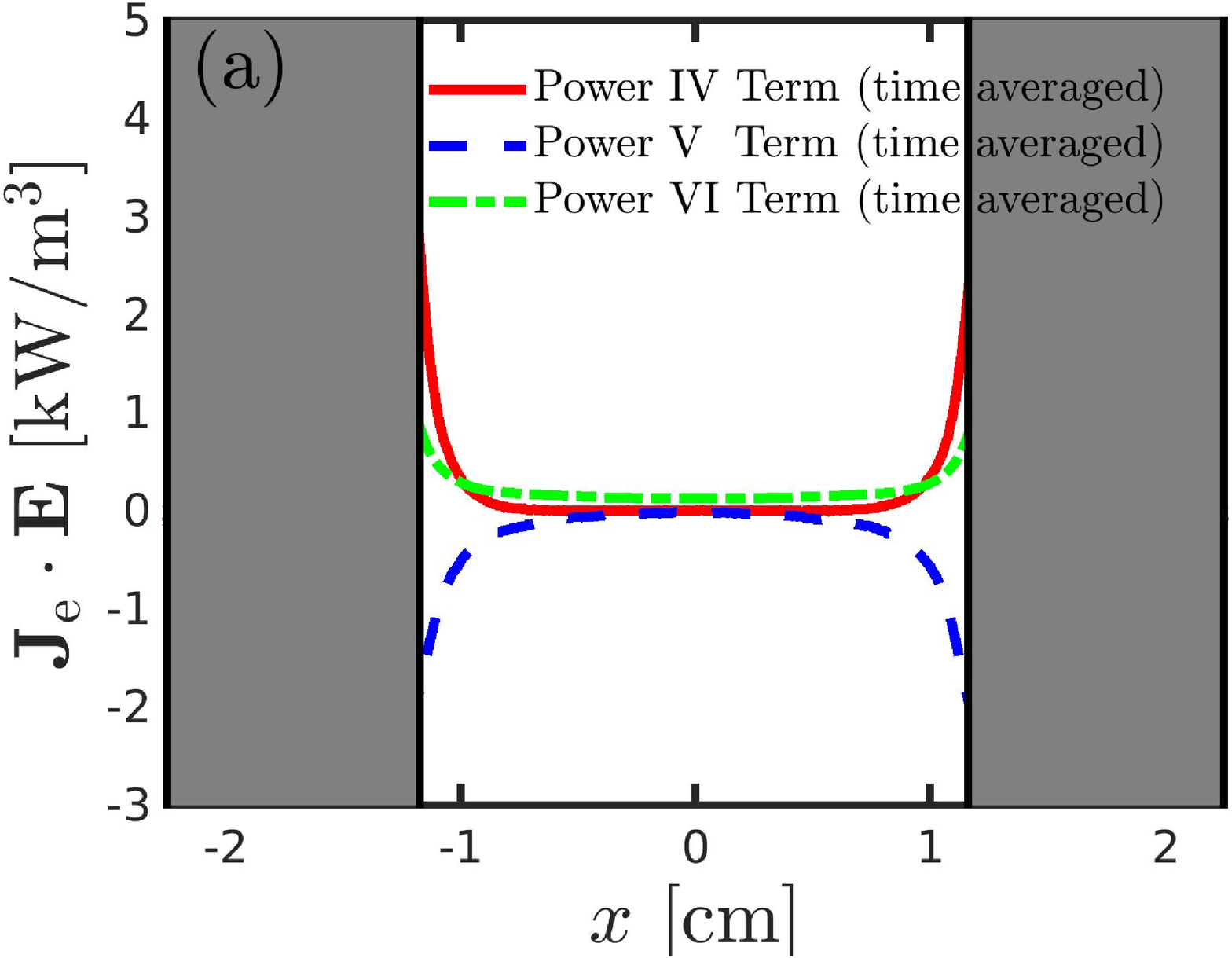}   
\includegraphics[width=8 cm]{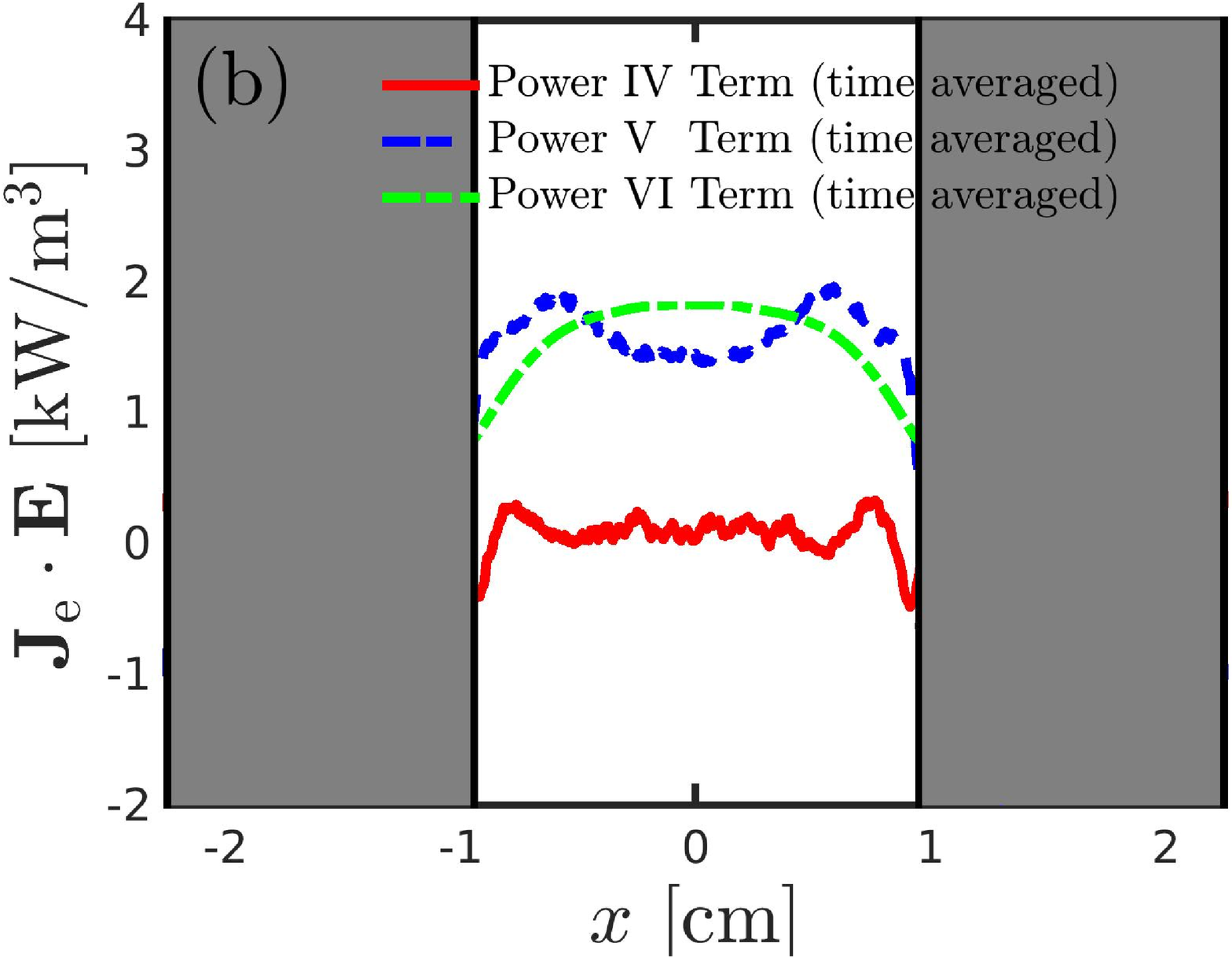}   
\end{center}
\caption{\label{fig9}The time averaged electron power absorption profile of Term IV (red line), Term V (blue dashed line), Term VI (green dotted dashed line) from Eq.~\eqref{equation5} from the left to the right (time averaged) sheath edge at (a) 100 mTorr and at (b) 10 mTorr for a parallel plate capacitively coupled oxygen discharge for 45 mm of gap separation driven by a 400 V voltage source at driving frequency of 13.56 MHz.}
\end{figure}

Figure \ref{fig9} (a) shows the time averaged electron power absorption profile at 100 mTorr of Term IV (red line), V (blue dashed line) and VI (green dotted dashed line) calculated using Eq.~\eqref{equation5} from the left to the right sheath edge.
All the three terms considered are flat and zero within the bulk gap length. We observe that Term IV (Term V) steeply increases (decreases) while approaching the sheath side of both the sheath edges. On the other hand, we observe that Term VI slightly increases while crossing both the sheath edges reaching small positive values on the sheath side of both the sheath edges. Finally, Figure \ref{fig9} shows that the main contributions to the time averaged electron power absorption at 100 mTorr comes from the pressure gradient related terms (Term IV and V) and from ohmic term (Term VI).

Figure \ref{fig9} (b) shows the time averaged electron power absorption profile at 10 mTorr of Term IV (red line), V (blue dashed line) and VI (green dotted dashed line) calculated using Eq.~\eqref{equation5} from the left to the right sheath edge. We observe that all the three terms considered have a parabolic behaviour over the bulk gap length.  Moreover, Term V is higher (lower) on the bulk side of both the sheath edges (within the inner electronegative core) than Term VI, while Term IV is sharply lower over the full gap length. In more detail, we see that Term IV is almost flat and zero within the discharge center up to the bulk side of both the sheath edges, and that it slightly increases while approaching both the sheath edges building equal local maxima. Then it steeply decreases while crossing both the sheath edges, building two almost equal local minima. On the other hand, Term V slightly increases while approaching the bulk side of both the sheath edges, building two almost equal local maxima, and it sharply decreases while approaching both the sheath edges. As regards to Term VI we see that it builds an absolute maximum at the discharge center and that it sharply decreases while approaching both the sheath edges.
\begin{figure}[!ht]
\begin{center}
\includegraphics[width=8 cm]{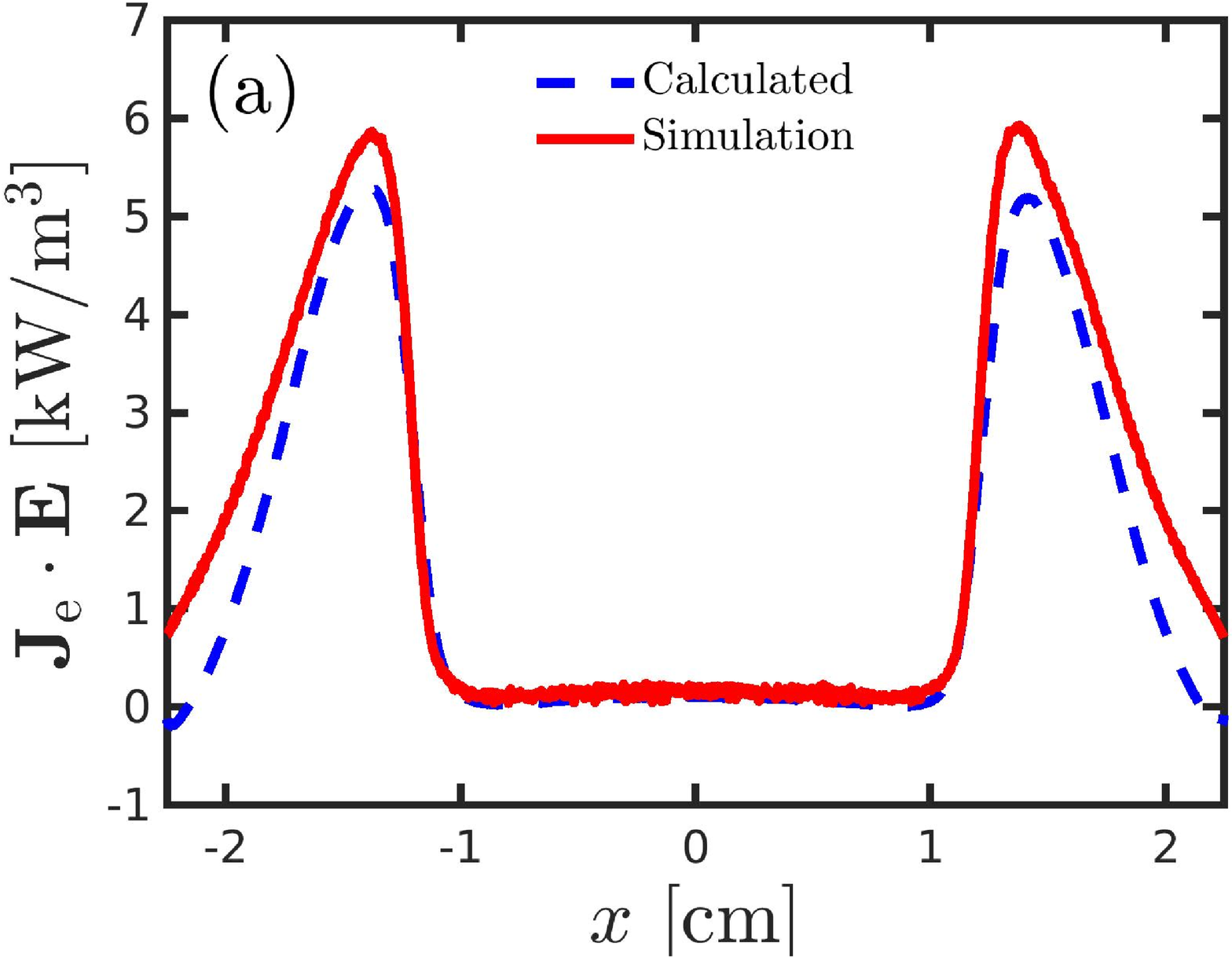}   
\includegraphics[width=8 cm]{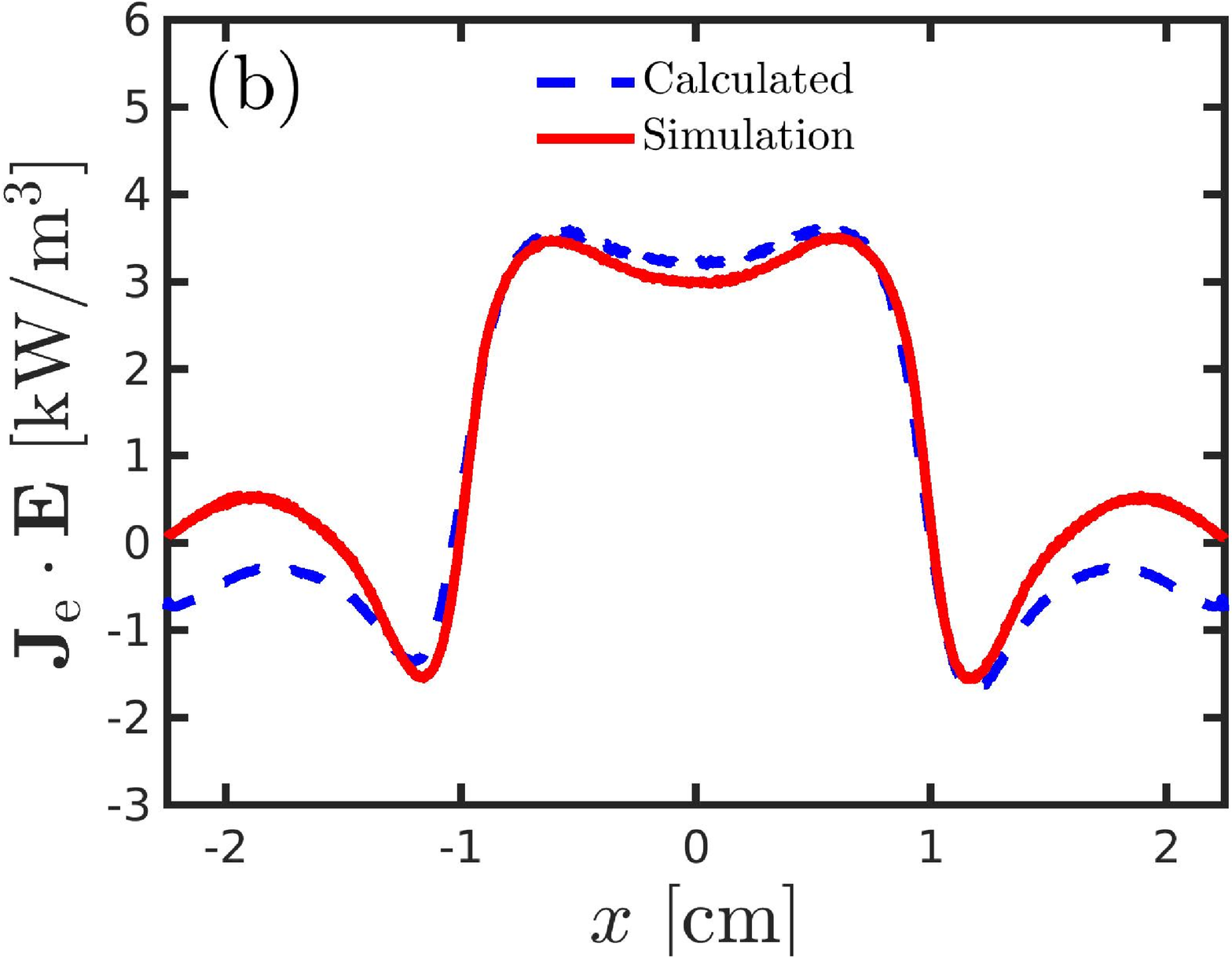}  
 \end{center}
\caption{\label{fig10}The time averaged electron power absorption profile calculated using Eq.~\eqref{equation5} (a) (blue dashed line) and the result from simulations (red line) over the full gap length at (a) 100 mTorr and at (b) 10 mTorr for a parallel plate capacitively coupled oxygen discharge for 45 mm of gap separation driven by a 400 V voltage source at driving frequency of 13.56 MHz.}
\end{figure}

Figure \ref{fig10} (a) shows the comparison between the time averaged electron power absorption calculated using Eq.~\eqref{equation5} (blue dashed line) and the result from simulation (red line) over the full gap length at 100 mTorr. In Figure \ref{fig10} (a) we see that the calculated time averaged electron power absorption overlaps almost perfectly with the result from the simulation over the bulk gap length up to the sheath side of both the sheath edges. Moreover, it slightly understimates the result from the simulation within the inner core of both the sheath regions up to the respective electrodes.

Figure \ref{fig10} (b) shows a comparison between the time averaged electron power absorption calculated using Eq.~\eqref{equation5} (blue dashed line) and the result from simulation (red line) over the full gap length at 10 mTorr. In Figure \ref{fig10} (b) we see that the calculated time averaged electron power absorption calculated using Eq.~\eqref{equation5} overlaps almost perfectly with the result from the simulation over the bulk gap length up to the sheath side of both the sheath edges. Moreover, the closer to the electrodes, the bigger is the difference between the calculation and the simulation.
\begin{figure}[!ht]
\begin{center}
\includegraphics[width=8 cm]{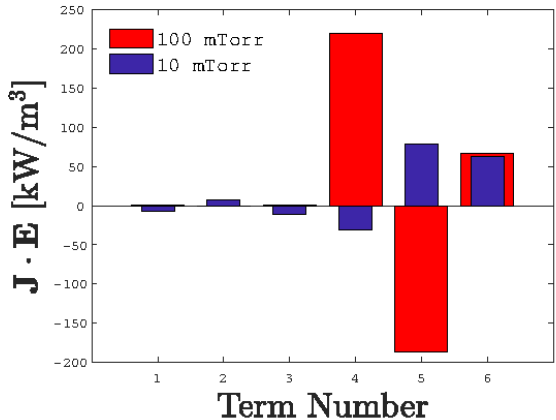} 
\end{center}  
\caption{\label{fig11}The space-time averaged electron power absorption profile terms calculated using Eq.~\eqref{equation5} at 10 mTorr (blue bars) and 100 mTorr (red bars) for a parallel plate capacitively coupled oxygen discharge for 45 mm of gap separation driven by a 400 V voltage source at driving frequency of 13.56 MHz.}
\end{figure}

Figure \ref{fig11} shows a comparison between the space-time averaged electron power absorption within the plasma bulk at 10 mTorr calculated using Eq.~\eqref{equation5} (blue bars) and at 100 mTorr calculated using Eq.~\eqref{equation5} (red bars). The space average has been taken within the bulk region only using the time averaged sheath locations as already discussed at the beginning of Section \ref{RaD}. The histogram has been constructed considering the calculated space-time averaged electron absorbed power and then building the following quantity
\begin{align}
\left(J_{\rm e} \cdot E\right)_{X \ \text{Term Percentage}}= \frac{ 100 \times \left(J_{\rm e} \cdot E\right)_{X \ \text{Term}}}{\left(J_{\rm e} \cdot E \right)}
\end{align}
where $\left(J_{\rm e} \cdot E \right)_{X \ \text{Term}}$ labels the space-time averaged electron power absorption related to the $X$ term, where $X$ refers to term I, II ... VI in Eq.~\eqref{equation5}. In Fig.~\ref{fig11} we observe that in the 100 mTorr case the space-time averaged electron power absorption comes from the pressure terms (Term IV and V) and the ohmic Term (Term VI). Moreover, we see that Term IV is positive while Term V is negative and they share almost the same magnitude in the absolute value, while Term VI is the smallest. Therefore, the ambipolar term is a power absorption term while the electron temperature gradient presents power loss (electron cooling). The main  electron power absorption at 100 mTorr is due to the pressure gradient terms. At 10 mTorr the situation has changed drastically. First of all Term IV and Term V flip their sign with respect to the 100 mTorr case and are sharply smaller in the absolute value compared to the 100 mTorr case. In this context the ohmic term's magnitude (Term VI) has not significantly changed compared to the 100 mTorr case but now shares the same order of magnitude with respect to both Term IV and Term V. Moreover, in Figure \ref{fig11} we recognize a general pattern where the ohmic term (pressure term) contribution in the space-time averaged total electron power absorption increases (decreases) when the the total pressure decreases (increases). Such a behaviour is expected and has also been observed by \citet{vass20:025019}. The only difference is that \citet{vass20:025019} find the pressure term contribution in the space-time averaged electron power absorption to be negligible at low pressure. Like \citet{vass20:025019} we find the ohmic power absorption to be important even at low pressure. Finally, at 10 mTorr we observe the presence of small contributions coming from Term I, II and III respectively.The inertia terms I and III provide electron cooling while the electron density gradient term II contributes to electron power absorption.
\section{Conclusion}\label{Conlc}
The one-dimensional object-oriented particle-in-cell Monte Carlo collision code oopd1 was applied to explore the properties of the electric field and the electron power absorption at different time steps and time averaged over a full rf cycle within the plasma bulk in a capacitively oxygen coupled discharge at both 100 and 10 mTorr for 45 mm gap distance.  At both 100 and 10 mTorr the fluid model presented by \citet{schulze18:055010} was applied.

At 100 mTorr at both $t / \tau_{\rm rf}=0.25$ and $t / \tau_{\rm rf}=0.75$ the main contributions to both the electric field and the electron power absorption are due to the electron inertia term related to the temporal gradient of the electron velocity (Term I), the gradient pressure related terms (Term IV and V) and the ohmic heating term.
At $t / \tau_{\rm rf}=0.50$ the main contributions to both the electric field and the electron power absorption come from the pressure gradient related terms (Term IV and V) and from the ohmic heating term (Term VI). We have also shown that the pressure gradient related terms and the ohmic term contribute to the time averaged electron power absorption, while only the pressure gradient related terms contribute to the time averaged electric field.

At 10 mTorr at $t / \tau_{\rm rf}=0.25$ and  $t / \tau_{\rm rf}=0.75$ the main contributions to both the electric field and the electron power absorption come from the pressure gradient related terms (Term IV and V) only, while at $t / \tau_{\rm rf}=0.50$ a small additional but not negligible contribution from the drift field (Term VI) has been observed. Moreover, in the time averaged case, the main contributions to the electron power absorption come from both the drift field (Term VI) and the pressure gradient related terms (Term IV and V).  We have also shown that the pressure gradient related terms and the ohmic term contribute to the time averaged electron power absorption.
 
\
 
\acknowledgements

This work was partially supported by the  Icelandic Research Fund Grant
No.~163086 and  the University of Iceland Research Fund.  



\section*{Data Availability}
The data that support the findings of this study are available from the corresponding author
upon reasonable request.

%
%

%
%

\end{document}